# Mistaken Identity and Mirror Images: Albert and Carl Einstein, Leiden and Berlin, Relativity and Revolution

Jeroen van Dongen

*Institute for History and Foundations of Science & Descartes Centre, Utrecht University, the Netherlands*
*Einstein Papers Project, California Institute of Technology, Pasadena CA 91125, USA*

Albert Einstein accepted a "special" visiting professorship at the University of Leiden in the Netherlands in February 1920. Although his appointment should have been a mere formality, it took until October of that year before Einstein could occupy his special chair. Why the delay? The explanation involves a case of mistaken identity with Carl Einstein, Dadaist art, and a particular Dutch fear of revolutions. But what revolution was one afraid of? The story of Einstein's Leiden chair throws new light on the reception of relativity and its creator in the Netherlands and in Germany.

## Introduction: Preparations in Leiden

On December 21, 1919, Hendrik Antoon Lorentz offered Albert Einstein a special position: He was asked if he would like to join the University of Leiden as a *bijzonder hoogleraar* or, in Lorentz's words, as a "let's say 'special professor.'" The Leiden University Fund (LUF), a private fund supported by donations from alumni and students, would endow the chair, and Einstein would be expected to travel to Leiden only once or twice a year; to hold classes or seminars would be appreciated, but not required.[1]

Einstein liked Lorentz's proposal very much, as he informed his close friend Paul Ehrenfest, who would be his colleague in Leiden. Ehrenfest, in turn, was as "excited as a child" about the "Einstein present" that the Leiden professors were preparing for themselves. Einstein soon started thinking about his inaugural lecture: It would have to be a lecture on the ether, since Lorentz had asked him to discuss publicly his latest views on its existence. In kind consideration of Lorentz's ideas, but also as a reflection of his latest reassessment of the Mach principle, Einstein now found that all along, instead of claiming the nonexistence of the ether,[a] he only should have argued for the "non-reality of the ether velocity."[2] In any case, Einstein wrote to Lorentz on January 12, 1920, that he would gladly accept the appointment, as he had been presented with a truly "fairytale-like" offer.[3] Leiden's University Council, that is, the LUF's Board, consisting of a large group of alumni,[4] formally approved the position on February 9, so the only remaining step was an official authorization by the Dutch government. But that, Lorentz assured Einstein, was to be expected "in a few weeks," and could be "solidly counted on."[5]

---

[a] In his inaugural lecture, Einstein discussed the metric field as a kind of ether, acting on, and being acted upon by matter; a definite state of motion, however, could not be ascribed to it. In deference to Lorentz, he even said that "the ether of the general theory of relativity is the outcome of the Lorentzian ether, through relativization"; see Einstein, *Äther* (ref. 2), p. 13; 178.

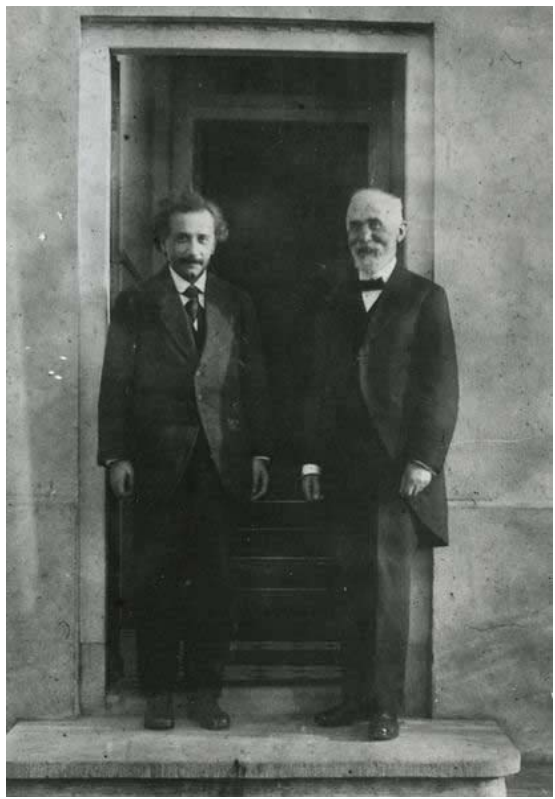

Figure 1: Albert Einstein (1879-1955) and Hendrik A. Lorentz (1853-1928) in front of Ehrenfest's house in Leiden in 1921; photograph by Paul Ehrenfest. Source: Boerhaave Museum, Leiden.

However, only after a long and halting process would Einstein's position be confirmed by the Dutch government—on September 21, 1920, well over seven months later. Only then would Queen Wilhelmina finally sign the decree that appointed Einstein to his Leiden chair. Why was there such an unexpected and long delay? What were the particular circumstances and concerns that led the Dutch government to take more than seven months to carry through on something that was, by all appearances, a trivial formality? I reconstruct below the factors that led to this delay,[6] and I discuss the broader political and socio-cultural contexts and concerns that motivated Einstein's appointment and undergirded the complications it encountered. It is a story with unusual twists and turns that shed light on the broader perception of Albert Einstein and of the reception of relativity, both in the Netherlands and in Germany.

## Why Einstein? The Internationalist Perspective

Paul Ehrenfest and Hendrik Lorentz first met Einstein a little less than a decade before Lorentz offered him the special Leiden chair. Ever since then, both felt a close personal bond with Einstein. This clearly was one of the reasons why they wanted to tie Einstein to Leiden. Lorentz and Einstein (figure 1) liked each other enormously, and were in awe of each other's scientific achievements; Lorentz was as a fatherly friend to Einstein, even if some physical distance seems to have been a necessary condition for their relationship.[7] Ehrenfest had a strong need to have Einstein near him. Both were German-speaking Jews of the same age, and their personalities, temperaments, and scientific interests resonated strongly. An intimate friendship, which included the families of both, had been gradually cemented after they first met in 1912. As Einstein wrote to Ehrenfest, "I have well learned to feel you as a part of me, and myself as a part of you."[8]

Having Einstein join the Leiden faculty, even if only as a part-time "special" professor, obviously would also greatly benefit physics there, and in the Netherlands in general, as Lorentz expressed.[9] In particular, Heike Kamerlingh Onnes very much looked forward to the opportunity to regularly discuss with Einstein the problems he faced in his cryogenic laboratory.[10] Kamerlingh Onnes had been less enthusiastic about the prospect of attracting Peter Debye, whose name had also circulated recently for a theoretical professorship in Leiden: Given Debye's stronger experimental interests, Kamerlingh Onnes had feared the loss of his uncontested stewardship of his laboratory. Ehrenfest had initially lobbied for an offer to Debye, being afraid that he himself would not be capable of maintaining the high quality of Leiden's theoretical physics; his efforts to tie Einstein to Leiden were inspired just as much by his sense of his own insufficiency.[11]

Einstein had been proposed for professorships in Leiden earlier. He had been offered Lorentz's chair in 1912, but at that time he had already committed himself to the Federal Institute of Technology (*Eidgenössische Technische Hochschule*, ETH) in Zurich; he had also been too intimidated by the prospect of having to follow in Lorentz's footsteps to accept.[12] Paul Ehrenfest eventually succeeded Lorentz, and Einstein, after a period in Zurich, moved to Berlin in 1914, where he combined a position at the Prussian Academy with the directorship of the newly-established Kaiser Wilhelm Institute for Physics and a professorship at the University of Berlin.

With the offer of a special chair in 1920, the Leiden physicists were careful to point out that they did not intend to lure Einstein away from Berlin, where life had become difficult after World War I. Earlier, in September 1919, Einstein had turned down yet another suggestion to join Leiden for a full professorship; on that occasion he had promised Max Planck that he would "not turn my back on Berlin, unless conditions prevail that would make such a move natural and correct."[13] Einstein, however, was in dire financial straits as he had to support his first wife Mileva Marić and their two children who were living in expensive Switzerland at a time when the German mark was quickly losing value. He had made clear to Ehrenfest that great sacrifices were being made to keep him in Germany, sacrifices he felt should not be snubbed by an ungrateful departure. Furthermore, his political wishes, entailing a parliamentary democracy and an end to the Wilhelminian "religion of power," had at last been realized following the collapse of the Imperial order. In view of these circumstances, Einstein believed that leaving Germany would be disloyal.[14]

The offer of a special, part-time professorship, however, he could accept.  For Einstein, the most important reason for welding a formal tie to Leiden was his close friendship with Ehrenfest and Lorentz.  He obviously also might have appreciated that the position came with a salary that, at 2000 guilders per year (the maximum annual salary of a full professor was 7500 guilders), was both substantial and was paid in a hard currency.[15] A further great attraction was that physics in Leiden was first-rate. Indeed, as Ehrenfest had already pointed out in September 1919: "Imagine what *human*-scientific environment you would have here: Lorentz, De Sitter, Onnes …, Kuenen, me and my wife, Droste, de Haas and his wife, Burgers, Zeeman, and all the time excellent young people, for example, Burgers and Kramers."[16] Exciting guests, such as Gunnar Nordström or Niels Bohr,[b] regularly passed

---

[b] Hendrik Antoon Lorentz (1853-1928); Heike Kamerlingh Onnes (1853-1926); Willem de Sitter (1872-1934), Professor of Astronomy; Johannes P. Kuenen (1866-1922), Professor of Physics; Johannes Droste (1886-1963), Lecturer in Mathematics; Wander de Haas (1878-1960), Professor of Physics at the Technical University of Delft; Geertruida de Haas-Lorentz (1885-1973); Johannes Burgers (1895-1981), Professor of Aerodynamics and Hydrodynamics in Delft; Pieter Zeeman (1865-1943), Professor of Physics at the University of Amsterdam; Hendrik A. Kramers (1894-1952), Assistant to Niels Bohr at the University of Copenhagen; Gunnar Nordström

through, and would often stay at the Ehrenfest home. Indeed, Einstein had only one reservation: he was concerned that any lectures he might give would add too little to the program already presented by the Leiden faculty. Yet, in January 1920 Einstein gratefully agreed to the appointment.[17]

Thus, there were abundant personal and scientific reasons for establishing a tie between Einstein and Leiden. Einstein would feel strongly at home in Holland, as may be evident from the expressive declaration he issued after the German invasion of the Netherlands in 1940:

> Holland has enriched humanity by its accomplishments of the mind and the arts, and it has contributed to the inner and political liberation of the individual by its struggle for justice. One is gratified by Holland's humane atmosphere as it brings together tolerance, broad understanding, humour and true helpfulness.[18]

The tenor of Einstein's declaration reflected both his outrage over the German invasion and his broader appreciation of Dutch culture in the interwar years (likely augmented by Einstein's sympathy for Spinoza and his era). A large part of his appreciation undoubtedly was due to the political ideals and efforts of some of Leiden's academic elite--who had been involved in securing Einstein's special chair.

The creation of Einstein's special chair followed closely after the conclusion of the Great War. Einstein had also been greatly offended by the outbreak of that war, which, in fact, had turned him into a committed pacifist. He had joined the "New Fatherland Association" in Berlin, a society that labored for peace during the conflict, and after the war he immediately sought to improve relations between the formerly warring nations. He wanted to reestablish ties between European intellectuals, and undertook multiple efforts to further this goal.[19] For instance, he welcomed French pacifist Paul Colin in Berlin, joined a private German committee to research war crimes committed in Lille and, in 1922, made a highly publicized visit to Paris.

Such efforts would resonate strongly with the ideals and actions of his Leiden colleagues. Of central importance for organizing Einstein's position were Lorentz and Cornelis van Vollenhoven (1874-1933, figure 2), Professor of Dutch East Indian Law and a prominent member of the LUF Board. Lorentz and van Vollenhoven also were leading figures in the Dutch Academy of Sciences: Lorentz as Secretary of its Section of the Sciences and van Vollenhoven, after 1920, as Secretary of its Section of the Humanities. Together they would spearhead Dutch efforts aimed at international reconciliation in the academic world.[20] Appointing Einstein would fit in well with these efforts or, as Paul Ehrenfest put it to him:

> *In fact* I believe that you, by spending a few weeks here, will contribute enormously, in an undemonstrative but therefore all the more powerful way, to the reestablishment of many disrupted scientific relations.[21]

Facilitating and negotiating international reconciliation was a core element of the professional identity of both Lorentz and van Vollenhoven; in the case of the latter, it also was an essential part of his scholarship. Van Vollenhoven saw a need to reorganize and systematize what he saw as the confusing patchwork of international law: Already in 1910, he was convinced that the Netherlands, as a small nation with an internationalist outlook,

---

(1881-1923), Professor of Mechanics in Helsinki, and Niels Bohr (1885-1962), Professor of Physics at the University of Copenhagen.

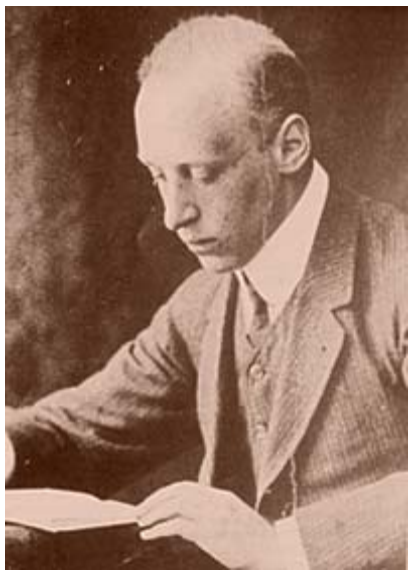

Figure 2: Cornelis van Vollenhoven (1874-1933). Source: Leiden University, Department of Law.

should take a prominent leading role in efforts to establish international peace. This, he believed, also would provide an opportunity to further the country's own plight and position.[22] He was convinced that progress toward international peace would be possible only if a supervisory and regulatory supranational police force were created to enforce the authority of an appropriate international court.[c] A second essential aspect of van Vollenhoven's role in Leiden was his great affection for its university, which he would serve as Rector and, among many other functions, as Secretary of the LUF. Under his stewardship, the LUF's endowment would grow substantially.[23]

On July 28, 1919, the International Research Council (IRC) and, a few months earlier, the *Union Académique de Recherches et de Publications*, its sister organization for the humanities, were created. Both organizations had barred from their memberships the academies of the former Central Power countries.[24] The neutral Dutch had committed themselves to lobby for the admission of the excluded countries; when asked to become a member of the IRC, the Dutch Academy, after heated internal debate, had decided to join but reserved for itself and its members the right to continue to collaborate with scientists from Germany and its former allies. Furthermore, it stipulated that its expectation was that united efforts eventually would be directed at reconciliation and the admission of the excluded parties to the IRC.

Just as the French IRC representative Émile Picard had feared, the Dutch Academy, under the stewardship of Lorentz and van Vollenhoven, soon began to act as an international mediator after joining the IRC.[25] For Lorentz such a role was a continuation of his efforts during the war, inspired by van Vollenhoven's ideas. His efforts were mirrored by those of Einstein, even if they were grounded in different ideals: Lorentz saw science as a national institution, and his attempts at international reconciliation were part of attempts to further the standing of Dutch science, while Einstein longed for the recovery of an international

---

[c] Van Vollenhoven perhaps influenced Einstein's thinking on international peace and disarmament, as van Vollenhoven's ideas seem to be reflected in Einstein's later positions during the Cold War; see, for example, Einstein's fall 1945 interview, "On the Atomic Bomb," in Rowe and Schulmann, *Einstein on Politics* (ref. 14), pp. 373-378, and Paul Doty, "Einstein and International Security," in Gerald Holton and Yehuda Elkana, ed., *Albert Einstein Historical and Cultural Perspectives: The Centennial Symposium in Jerusalem* (Princeton: Princeton University Press, 1982), pp. 347-367.

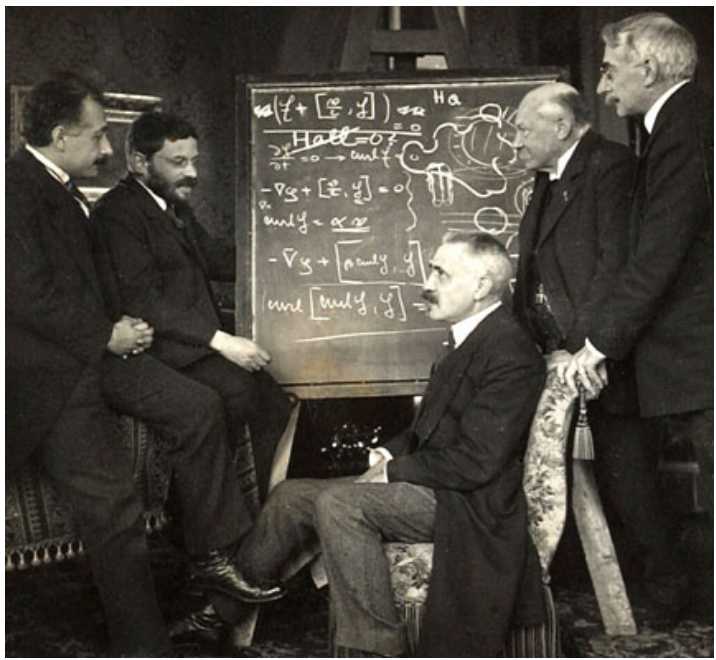

Figure 3: *Magnetwoche* in Leiden, 1920, left to right: Einstein, Paul Ehrenfest, Paul Langevin, Heike Kamerlingh Onnes and Pierre Weiss discussing the physics of paramagnets and superconductivity at Kamerling Onnes's home in Leiden. Source: Wikimedia.

community of scholars that would serve as an inspiration against pernicious nationalist sentiments.[26]

It thus seems clear that Lorentz and van Vollenhoven would see Einstein's appointment as an opportunity to assist their reconciliation efforts. When the procedure finally came to a close in October 1920, Einstein attended a small workshop in Leiden on the properties of paramagnets at low temperatures and the physics of superconductivity that had been organized by Ehrenfest and Kamerlingh Onnes, and was held under the auspices of the *Institut International du Froid*, with which the latter had close ties.[27] Based in Paris, the *Institut* aimed to promote research at low temperatures and to facilitate international collaborations; Axis countries, however, again had been excluded from membership. Einstein's participation thus carried strong symbolic value, in particular because the *Magnetwoche* was also attended by Frenchmen Pierre Weiss and Paul Langevin.[28] A picture taken at the event (figure 3) suggests cordial and productive discussions, regardless of nationality; Leiden and the Netherlands could once more be seen as facilitating the exchange of ideas and the reestablishment of personal relationships. In fact, Ehrenfest, earlier in 1919, when expressing his excitement about Einstein's future Leiden position, already had told Einstein that he thought the international position of Kamerlingh Onnes's cryogenic laboratory offered excellent opportunities in this regard, since visitors from abroad spent both long and short periods there for experimentation.[29]

Why were so many Dutch intellectuals keen to present themselves as international peace brokers? What motivated this peculiarly Dutch position? The Netherlands had maintained a policy of strict neutrality throughout the nineteenth and early twentieth centuries; it was, after all, a small nation hemmed in between the great European powers with substantial but vulnerable colonial interests.[30] German unification, and in particular Prussian dominance in the Empire, had led some late-nineteenth-century Dutch scholars to fear for their cultural and intellectual independence: They were afraid of an unduly large influence on the literature, practice, and institutions of Dutch science from an oversized neighbor to the east; in some disciplines (medicine in particular), there already were clear signs of such an undesirable dominance. Historians Willem Otterspeer and Joke Schuller tot Peursum-Meijer have observed that many Dutch academics, to allay their fears, began to emphasize a cosmopolitan spirit: fears of German dominance would be unfounded so long as the Netherlands retained its role as a mediator between the dominant cultures of Europe.[31] Seen in this light, the wake of the Great War actually presented an opportunity for Dutch scholars: By taking care of the international plight of German science, the Dutch could assert their desired independence. Appointing Einstein at this point in time then would confirm Dutch academics in their self-chosen role, and assure them of their international relevance.

**Einstein and the Politics of Dutch Appointment Practices**

On March 10, 1920, a month after its submittal, Paul Ehrenfest reported to Einstein that his appointment was still "crawling around in government offices." Nevertheless, Ehrenfest expected a quick resolution of the matter. He proposed to Einstein that he travel to Holland in the second half of April and hold his inaugural lecture on May 5. Einstein therefore finished work on his lecture in early April and scheduled a trip to Leiden for around May 1. To save money, at the urging of Ehrenfest, he had decided to have his lecture printed in Germany instead of the Netherlands. Einstein was not sure how to compose its title page, however, nor did he know how to include the customary dedications to the chair's curators and other faculty members. He wrote that he felt like a child that was about to experience its first day in school; he was particularly uncertain concerning the correct pronounciation of *Ik heb gezegt* (he spelled it incorrectly--it should be *gezegd*), which simply means "I have said" and is the traditional closing of Dutch inaugural lectures.[32]

Ehrenfest replied in mid-April with detailed instructions for the title page ("The Dutch title will necessarily have to be as follows: Down with the ether superstition!!—Lecture delivered at the assumption of the duties of special professor at the University of Leiden by A. Einstein"), and for the customary and elaborate acknowledgments. He decided for Einstein that he should pronounce *Dixi* instead of *Ik heb gezegd*, undoubtedly to Einstein's relief. But Ehrenfest still complained about the slow pace with which the appointment was being treated by the Dutch bureaucracy.[33]

The delays were not due to officials in Leiden: Immediately after the formal approval of the chair by the Leiden University Fund, van Vollenhoven approached the Queen, who ultimately held the proper authority to establish a chair and appoint Einstein. Her cabinet forwarded the request to the Ministry of Education, Arts, and Sciences, which in turn consulted its Board of Supervision ("Commissie van Toezicht, bedoeld in art. 201 der H.O. wet"). The Board saw no objections to allow the LUF to create a chair in physics. But it did raise a concern regarding the prospect of having Albert Einstein occupy that chair: objections possibly might be raised against his appointment "in view of his political principles."[34] To justify its concern, the Board referred the Minister of Education, Johannes Theodoor de Visser (1857-1932), to a recent newspaper article in the liberal *Nieuwe Courant* entitled "Berlin University Life."[35]

This article described the general hostility toward the immediate postwar German Revolution of November 1918 that prevailed among Berlin university students. They had resumed their old nationalist traditions, which had been suspended when they first tested the waters after the revolution; now, however, the "nationalist-reactionary spirit dominated again," and no opportunity to hold a nationalist protest went unobserved. The Berlin professors, too, had not learned anything from recent events. Still, the *Nieuwe Courant*'s correspondent noted, a minority did understand and appreciate the events of November 1918, among whom was Einstein.

Reactionary Berlin students and professors especially singled out pacifists against whom they vented their anger: Recently, the Dutch paper informed its readers, Georg Nicolai's lectures had been so severely disrupted that he had been forced to suspend them. Nicolai was Extraordinary Professor (*ausserordentlicher Professor*) of Medicine and Physiology, who within days after the infamous militaristic "Manifesto of the 93" was issued in October 1914 had drafted a counter "Manifesto to the Europeans"; a little later he had fled Germany owing to the condemnation of his publications. Now, in 1920, after the disruptions of his classes, he was declared unfit for teaching altogether, since a university senate committee had found him guilty of committing treason during the war. Einstein circulated a statement in support of Nicolai,[36] only to find his own lecture disrupted three weeks later: "instead of appreciation for the fact that the famous man of the theory of relativity teaches at their university, [rightist students] try to make his classes impossible by obstruction," the *Nieuwe Courant* reported.

Apparently unknown to the Dutch correspondent was that the obstruction of Einstein's lecture had been motivated in part by his opening up of his classes essentially to anyone who wished to attend. Tuition-paying students objected, and had successfully protested Einstein's open-door policy to the university authorities. Einstein initially demurred, and preferred to refund all student fees. This action was taken, but only after the tumultuous lecture had taken place. Another context, however, was also relevant to this story: Some Berlin newspaper reports claimed that anti-Semitic catcalls had been made during the disruptions. Einstein and the Rector of the University denied this. Yet, it would be quite imaginable: By opening up his lectures, Einstein had wished to accommodate a large group of immigrant students that consisted mostly of poor Jews who had fled pogroms in Eastern Europe. In the end, however, he decided against his own initiative, believing that these students would be better served by classes at some institution other than the University of Berlin.[37] The Dutch authorities in The Hague were obviously unaware of these developments. In any case, they already had become alarmed by the article in the *Nieuwe Courant*: they decided that Einstein merited a more thorough look before a decision on his Leiden chair could be reached.

The Dutch government at the time consisted of a coalition of three popular confessional parties: the Roman Catholic State Party, the Christian Historical Union, and the protestant Anti-Revolutionary Party (whose name primarily reflected its opposition to the values of the French Revolution). The government had just created a new Ministry of Education, headed by de Visser, a former clergyman, and he and his Ministry were now dealing mostly with Einstein's case.

Officially, professorial searches in the Netherlands would not consider a candidate's political positions or religious beliefs; in reality, however, things worked differently.[38] Ever since a reform of higher education in 1815, university professors were regarded as such senior civil servants that their appointments required royal approval. After a major constitutional redraft in 1848 that limited the monarch's actual powers, this in effect meant that these appointments were reviewed by the central government in The Hague, which also decided on the memberships of the governing boards of the universities; these "curators" often were seasoned administrators who themselves had served in The Hague. In the end, this entailed that, formally, local professors had little clout in the choice of their future colleagues. Nevertheless, both professors and administrators usually had the same elitist background: religiously, they belonged to the well-established Dutch Reformed Church, and politically they were liberal; candidates proposed by the sitting faculty usually would have few problems being approved.

The popular confessional parties, the same parties that were in the coalition government that would review Einstein's appointment in 1920, wanted to change this *status quo*. At the beginning of the twentieth century, the protestant Anti-Revolutionary Party (ARP), whose constituents were members of the breakaway neo-Calvinist Reformed Church in the Netherlands, had labored to create special professorships at Dutch state universities. The ARP's goal was to enable private societies and foundations to endow chairs and to have them filled by candidates who were not members of the traditional elite; of course, the ARP wished to have its own constituency's candidates appointed to such special chairs, in particular in the humanities and law, and most definitely in theology. Leiden initially opposed this development, even if it offered the possibility to create chairs that were motivated primarily by scientific interests, or to broaden the curriculum offered to students. [39] Eventually, however, Leiden used the LUF to create special professorships, like Einstein's, for exactly these reasons. [d]

The new Ministry of Education was headed by Christian party politicians from 1918 until 1933. It actively considered the religious denomination of professorial candidates, much to the dismay of the sitting faculty. Its goal was to achieve more diversity, reflecting the religious diversity of Dutch society at large. Nonetheless, socialists--who were regarded as too extreme and a liability--and women still would not be considered for positions at Dutch state universities. Objections also could be raised in regard to German candidates:[40] local talent should be favored, since German chairholders often had chosen to return to Germany when opportunities arose, and, as noted earlier, Dutch academics generally were fearful of being overrun by the dominant culture next door.

Thus, in the end, the official policy of not considering a candidate's personal background was regularly sidestepped in backroom dealings. Nevertheless, the majority of appointees were still of familiar liberal pedigree, and the issue actually was considered to be of immediate relevance only if a chair in the humanities was involved, in particular a professorship in theology.[41] Given these circumstances, was it to be expected that Einstein would be considered a problematic candidate in The Hague? He was left-leaning and politically active: this would have raised discomfort and could prompt inquiries, as it indeed did following the article in the *Nieuwe Courant*. That Einstein was Jewish, should it have come up, would not have helped, nor that he could be seen as German. But there also were mitigating circumstances: In Einstein's case, the Ministry of Education was dealing with a safe chair in physics, and only a special chair at that, funded by a private organization. Thus, in the end, the Dutch civil service need not have been alarmed by a fear of inevitable and heated debates in parliament--yet, it soon would be alarmed, very much alarmed. Why?

## What Went Wrong in The Hague?

"Why is the confirmation of your appointment being held up for so long?" Paul Ehrenfest asked Einstein in a letter of April 13, 1920, two months after the LUF had sent its request to the Ministry of Education, and a full month after its Board had referred the Minister to the article in the *Nieuwe Courant*. Ehrenfest still expected the matter to be resolved promptly, however, and remained convinced that it was an excellent idea for Einstein to travel to Holland two weeks later.

---

[d] The LUF generally would spend its funds on improving education, for instance by awarding student and faculty scholarships. In 1918 it would endow a first special chair in public finances, held by Anton van Gijn, a former Minister of Finance; in 1920, along with Einstein's chair, the LUF funded a special chair in tropical medicine; see Otterspeer, *Een welbestierd budget* (ref. 4), p. 58. It was van Vollenhoven's idea for the LUF to endow special chairs; see de Kanter-Van Hettinga Tromp and Eyffinger, *Cornelis van Vollenhoven* (ref. 22), p. 31.

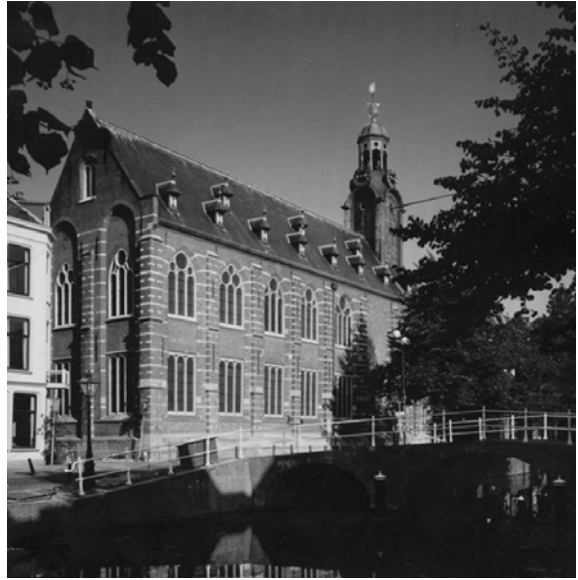

Figure 4: *Academiegebouw* at Leiden University.

Einstein arrived in Leiden on May 7. After a few days he reported home that his appointment was still running into "difficulties." He also indicated that the ground for those difficulties was his "hot political reputation," which would have made the Dutch authorities wary.[42] His wife, Elsa Einstein-Löwenthal, was unhappy about these tidings:

> So everywhere people have it in for you because of your socialist beliefs! Even in Holland! Do me a favor and don't act like such a furious Socialist; you are not one any more so than Ehrenfest and many others! Please finally put an end to this stupid talk; at last, you are regarded everywhere as a raging revolutionary…. It is already bad enough that you won't get the Nobel Prize because of this, and it should not go any further. A critical mind like you is not a communist![43]

Elsa, of course, was right: Einstein was not a communist. He thought the Bolsheviks do "not seem so bad," but found their theories "ridiculous." He was a democrat first, although at this time he was not convinced of the benefits of a welfare state; all in all, he seems not yet to have given too much thought to issues in political economy.[44] It is therefore surprising that his wife said that he had the reputation of a "raging revolutionary." Clearly, his reputation was much sharper than his actual positions warranted.

In any case, Einstein still could not deliver his much-anticipated inaugural lecture in May, because the approval of his chair had not yet materialized. When some weeks later he set out to return to Berlin without the appointment to his new special chair, his wife scolded: "Now where is the appointment to the Dutch professorship? You have been put in a ridiculous position." Elsa reminded him of the hurrying she had to impose on the staff at the Springer publishing house, which had printed Einstein's inaugural lecture, and she worried about who was going to pick up the bill for it now that it could not be circulated.[45] Instead of delivering his inaugural lecture, Einstein spoke to the Leiden Society for Scientific Lectures in the *Academiegebouw* (figure 4) with a number of dignitaries in attendance, and was elected as a member of the Royal Academy of Sciences in Amsterdam.[46] Despite these festive occasions, however, Einstein's Leiden colleagues were "greatly embarrassed."[47] As Lorentz informed him: "if it had been in our power, the whole matter would have been ready on 10 February,"[48] but even a last-minute attempt at intervention in The Hague by Kamerlingh Onnes yielded no result.

The Dutch authorities were indeed quite apprehensive about appointing Einstein. On March 26, 1920, more than a month before Einstein would travel to Holland, Minister of Education de Visser had invited LUF Board member van Vollenhoven and N.C. de Gijselaar, Mayor of Leiden and President-Curator of the University of Leiden, for a private consultation.[49] After the article in the *Nieuwe Courant* had appeared, some very disturbing notices had been brought to the attention of the Ministry of Education: the office of the Attorney General had informed Minister de Visser that a year earlier it had received information from the Justice Department that:

> [A] certain dr. Eisenstein [*sic*, meant was Einstein] would be sent to the Netherlands from the revolutionary side in Germany, to start up a propaganda service in this country. He would come here with a passport that would appear to be in order, but is presumably false, and made in Berlin.[50]

That was not all: Eisenstein's (Einstein's) partner, a countess named "von der Hagen," had turned up in France to participate in "bolshevist propaganda activities"; she too was suspected of traveling on a false passport.[51] The Attorney General further referred Minister de Visser to a distressing secret memorandum that had been drafted on June 18, 1919, by a military intelligence officer for the benefit of a regional command center in Breda, close to the Dutch-Belgian border:

> I was informed yesterday that Dr. Einstein [not Eisenstein] and Countess Olga von Hagen [not von der Hagen] live in Berlin, Wilmersdorf, Uhlandstrasse. As I have pointed out to you earlier, both have lived in Brussels during the occupation of Belgium, where Dr. Einstein has repeatedly attempted to promote a revolution among the Belgians, and Olga von Hagen often published under the pen name "The Red Countess." Her writings also intended to bring the Belgian population to revolt against the government. From receipts that I have seen lately, it seems to me that they maintained quite an extravagant lifestyle in Brussels. Both persons are closely watched and their departure for the Netherlands will certainly be reported in time.[52]

Minister de Visser (figure 5) must have been truly alarmed: this Einstein appeared to be a most undesirable character indeed, and certainly not someone he would like to appoint to a Leiden professorship, even if it was only a "special" chair.

But de Visser did not dismiss the LUF request outright; as noted, he had called in van Vollenhoven and de Gijselaar for consultation. The former reported back to de Visser the day after their meeting:

> It is with great pleasure that I can confirm to you that we are dealing with a case of mistaken identity. Professor Einstein has *never* lived at the address that you mentioned, is married to a little Jewess that has the same name as he does, does *not* consort with countesses, and has *not* lived in or spent time in Brussels during the war….[53]

Van Vollenhoven pointed out that there are many Einsteins living in Berlin (among them Alfred Einstein, the well-known musicologist) and that "our Albert E." was a pacifist and an opponent of Prussian warmongers, but that could hardly be held against him. He again urged

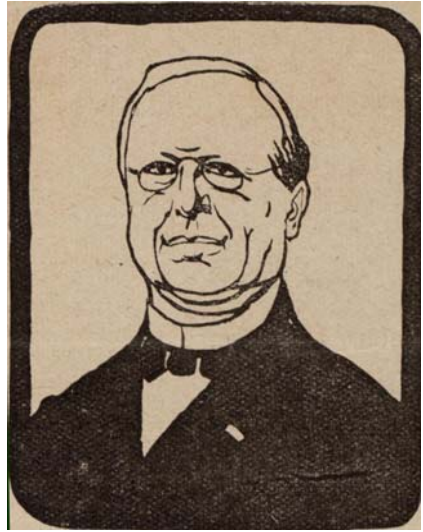

Figure 5: J. Th. de Visser (1857-1932), Minister of Education, Arts and Sciences in 1918-1925, by socialist artist Albert Hahn.

Minister de Visser to expedite the confirmation of Einstein's appointment, so that he could deliver his inaugural lecture shortly after the Easter break, sometime in April.

As we know, Einstein's appointment was not expedited. It appears that Minister de Visser and his staff simply did not do anything for another six weeks, until Einstein was about to arrive in Holland in early May. Then, instead of approving Einstein's appointment, de Visser called in another authority for assistance: He sent a request to his colleague, the Minister of Foreign Affairs, asking him if no objection could be raised against the appointment of Albert Einstein (and not "the musician Alfred Einstein");[54] additional reassurances from the physicists that Einstein "finds communism a stupidity" had been no help.[55]

The Ministry of Foreign Affairs took two weeks to reply: on May 20, it reported to Minister of Education de Visser:

> An inquiry, conducted by her Majesty's ambassador in Berlin, [has established that] Professor Alfred [*sic*] Einstein is part of the far left wing of the democratic party. The notion that he would be connected to revolutionary circles could not be confirmed to Baron Gevers. The problems that the named Professor had some time ago with his students would not have been the consequence of his political views, but should be seen entirely as a result of the wish of the students to protest against the decision of Professor Einstein to allow the general public admission to his lectures. He would have been brought to this decision owing to his great vanity.[56]

So Baron Gevers, the Dutch Ambassador in Berlin, established that even if this Alfred or Albert Einstein was not the safest of choices, he was no Bolshevik rabble-rouser. De Visser, finally reassured, then dutifully passed on his approval of the LUF's request to create a special chair in physics in Leiden to the full Council of Ministers and the Cabinet of the Queen. The Queen's officials replied on June 9, confirming the approval of the creation of the special chair.[57] But by this time, of course, Einstein had already left the country.

However, even if Einstein had stayed in the Netherlands a little longer, there still would not have been enough time for him to accept the chair. Pushing papers from one high office to another took another couple of weeks, and only on June 25 did Queen Wilhelmina actually sign the official decree that allowed the LUF to create the chair.[58] Nonetheless, Einstein at this point *still* could not be appointed to his special chair. Yet another obstacle had arisen: Einstein did not hold a Dutch doctorate. His appointment therefore needed one more full round of ministerial approvals and royal confirmations. The LUF thus again turned to Minister de Visser on July 26, 1920, and humbly submitted Einstein's *curriculum vitae* for the perusal of his officials.[59]

De Visser again dutifully passed the request on to his Board of Supervision, once more seeking approval of the candidate Albert Einstein. This time the Board could "advise favorably,"[60] but only on the condition that de Visser was absolutely sure that:

> [All] doubts have been eradicated regarding the identity of the appointed professor and Dr. Einstein, who during 1919 has been observed as decidedly committed to the communist principles.

To erase any lingering doubts, to be absolutely certain, once and for all, and to inform Minister de Visser as best it could, the Board included a physical description of "the communist Dr. Einstein":

> [R]ound head; robust physical appearance; 1.72 m[eters] tall; baldness at the temples, and advancing beyond the middle of the forehead; sideburns halfway along the ears; no moustache or beard; straight nose; big mouth; eyes stand apart from each other and are of an indefinable grey; wears a lorgnette or glasses with horn rims, sometimes also a monocle; on his right hand a ring with a blue gem.

This indeed would clear up all possible confusion, once and for all, since Albert Einstein obviously did not fit this description. Thus, on September 16, 1920, Minister de Visser finally requested Queen Wilhelmina to issue a decree that appointed Einstein to the newly-created special chair in Leiden. She did, five days later: Einstein now at last could hold his inaugural lecture on the "Ether and the Theory of Relativity" at 2 P.M. on October 27 in the large auditorium of the University of Leiden.

**Troubles in Berlin**

The almost five months between Einstein's departure from Leiden at the end of May 1920 and his return on October 21 had quite difficult for him--but his difficulties did not have much to do with the Dutch delays: Berlin had seen a full-blown anti-Einstein campaign kicked off by a certain Paul Weyland (1888-1972, figure 6).[61] Weyland's campaign constituted the first apogee of what has become known as the anti-relativity movement. Although Weyland's actions appear to be unrelated to the Dutch delays, understanding the latter will nevertheless assist us in opening up a new perspective on the anti-relativists, as we will soon see.

On August 6, 1920, Weyland published an aggressive newspaper article claiming that Einstein, among other things, had plagiarized the work of others when he had formulated the theory of relativity. The article further contained thinly veiled anti-Semitic allusions and concluded that "German Science" had to close ranks and "settle scores."[62] A few days later, Weyland announced a series of lectures that would reveal the truth about Einstein, who had unduly captivated the public's imagination ever since the announcement of the eclipse results in 1919. Now, however, twenty lectures, to take place in Berlin's Philharmonic Hall and organized by the Working Society of German Scientists for the Preservation of Pure Science, were going to break that spell.

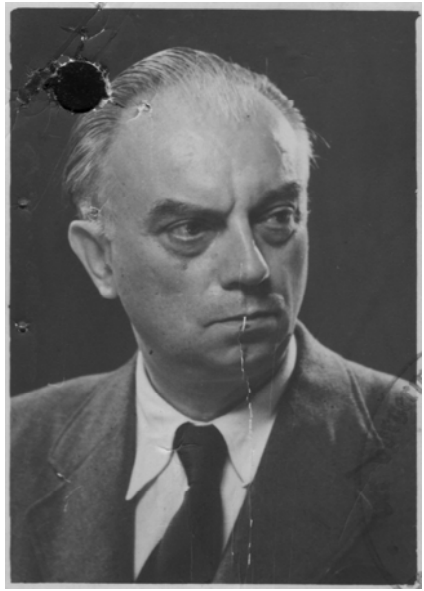

Figure 6: Paul Weyland; Source: Andreas Kleinert, private copy.

In all likelihood, the Working Society had only one member who had no scientific qualifications whatsoever: Weyland himself.[63] Nevertheless, owing to the hubbub he had created, the large auditorium of the Berlin Philharmonic was filled to capacity on August 21, 1920. There were two speakers: Weyland and Ernst Gehrcke,[64] an experimental physicist who had publicly resisted relativity as early as 1911. Gehrcke reportedly presented his arguments with some calm, but Weyland ranted and raved like a demagogue.[65]

> *Meine Damen und Herren!* Hardly ever in science has a scientific system been set up with such a display of propaganda as the general principle of relativity, which on closer inspection turns out to be in the greatest need of proof![66]

Weyland claimed that Einstein had resorted to propaganda tactics because he had been unable to counter his academic critics in any other way. His charge of propaganda and media hyping was a familiar strategy among Berlin anti-Semites, fueled by the circumstance that some newspapers, for instance, the liberal and widely circulated *Berliner Tageblatt*, were owned or edited by Jews.[67]

Paul Weyland was an obscure product of postwar Berlin, and following his action against Einstein (which he would exploit for a number of years) he gradually faded into the margins again.[68] Einstein, like Weyland himself, initially believed that the latter had the support of a fair number of physicists and philosophers--these, like Nobelist and prominent anti-relativist Philipp Lenard, however, quickly melted away owing to the controversy that the Berlin event produced. After Weyland's and Gehrcke's lectures, only one other lecture of the announced series actually took place, and many anti-relativists soon distanced themselves from Weyland. Nevertheless, they would continue their--often highly vocal--opposition to the theory of relativity for many years.

Weyland appears to have been motivated by an opportunistic desire for the limelight. His criticism was however foremost inspired by the many rightist political frustrations he bore. He was an active member of the *Deutschnationale Volkspartei*,[69] a nationalist-conservative party that strongly opposed the Revolution of November 1918 and the Weimar Republic. Culturally, he resisted modernist developments, and in an increasingly vehement manner he defended *deutsch-völkisch* values against perceived onslaughts from Poles and Jews.[70]

According to Weyland's assessment, which he presented to his audience at the Philharmonic, Germany was suffering from intellectual and moral decay that had attracted "all kinds of adventurers, not only in politics, but also in art and science." He claimed that the theory of relativity was being "thrown to the masses" in exactly the same way that the leftist "Dadaist gentlemen" promoted their products. These, too, had nothing to do with observation of nature, as little as had the theory of relativity. Now, finally, a movement had arisen that was going to fight this "scientific Dadaism."[71]

Einstein was in the audience during Weyland's and Gehrcke's lectures. He wrote a reply in the *Berliner Tageblatt* a few days later. Upset and angry, he was convinced that his opponents were primarily politically motivated ("if I were a German nationalist with or without a swastika instead of a Jew with liberal international views, then ...").[72] Einstein deemed the speakers in the Philharmonic "not worthy of an answer from my pen," and found that Lenard, even if he had done valuable experimental work, had "not produced anything" of lasting value as a theorist. The Dutch press, like the German press, closely covered the events in Berlin,[73] which was how Paul Ehrenfest learned of them: He immediately wrote to Einstein, imploring him not to allow himself to be dragged down into the dirt by his opponents. He also assured Einstein that, should he wish to leave Germany (as newspapers were reporting) Leiden undoubtedly would be able to accommodate him with a full position, even if initially it could commit itself formally for only a few years.[74]

When Einstein replied to Ehrenfest's letter, he had calmed down considerably: He was already convinced that the "anti-relativity company is pretty much broke," and that he would not have to leave Berlin.[75] Lorentz meanwhile had informed Einstein that the special professorship would now finally be fully approved, and that he could expect to hold his inaugural lecture at the end of October.[76] Before Einstein could leave Berlin for Leiden again, however, he first had to attend the meeting of the *Gesellschaft Deutscher Naturforscher und Ärzte* (Society of German Scientists and Physicians) in the resort town of Bad Nauheim. A public discussion on relativity had been scheduled there for September 23, and everyone, including Germany's newspapers, expected a standoff between Einstein and his opponents. Indeed, the audience, which had turned up in a substantial number, was treated to an intense yet factual exchange between Einstein and Lenard; Einstein would describe it as a "cockfight."[77] Both came away from the debate feeling frayed; in fact, Einstein's wife Elsa had fallen ill in Bad Nauheim, in part owing to the nervous and tense atmosphere surrounding the debate.[78] Einstein surely would have been glad that following the event, after a comforting sojourn with Elsa and his children to her hometown of Hechingen, he finally could travel once again to Holland to deliver his inaugural lecture.

## What Dr. Einstein?

But who was the mysterious "Dr. Einstein", with "baldness at the temples," "sideburns" but "no moustache or beard," "straight nose," "big mouth," eyes that "stand apart from each other and are of an indefinable grey," who "wears a lorgnette or glasses with horn rims," and most disturbing of all, was "committed to the communist principles"? Was there a communist Einstein in Berlin or elsewhere who actually fit this description?

Indeed there was. The railway police in Bamberg, Bavaria, reported on June 22, 1919, that they had "apprehended [a] person, presumably Einstein, ... about 165 cm tall ...; without

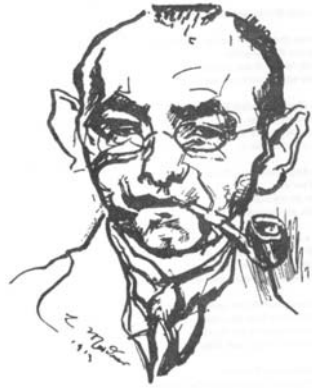

Figure 7: Carl Einstein (1885-1940), by Ludwig Meidner in 1913. Source: Ludwig
Meidner Archive, Jüdisches Museum der Stadt Frankfurt am Main.

beard, with sideburns, glasses, dark hair, bald on his forehead and the back of his head." He
also had "a deep scar from surgery behind his right ear" and "speaks slowly, calmly, and
without an accent."[79] He had been stopped on June 14 at the Bavarian border while traveling
on a train from Berlin to Nürnberg, and had identified himself with a false passport--a
military identification card in the name of a certain Paul Karl Körcher, which had greatly
alarmed the police officers: they believed they were facing none other than Max Levien, the
fugitive leader of the Munich branch of the German Communist Party.[80] Levien had been a
prominent foreman of the short-lived Bavarian Soviet Republic (April 7 – May 2-3, 1919),
and widely-circulated wanted posters had offered a reward of 30,000 marks for aid in his
arrest.  Levien's fellow Bavarian communist leaders had already been either killed or
arrested. Nonteheless, neither Levien nor Paul Körcher had been apprehended: the railway
police had indeed arrested Einstein--not Albert but Carl Einstein, author, avant-garde art
critic, and radical revolutionary.

Carl Einstein also was the person the Dutch authorities had confused with Albert
Einstein:  Carl Einstein did live in Berlin on Uhlandstrasse (at number 32, actually only a few
blocks away from Albert Einstein at Haberlandstrasse 5), and he indeed was "consorting"
with a countess, Aga von Hagen. But why would the Dutch authorities confuse a Berlin
communist with the world's most famous physicist?  The neutral Dutch had followed
developments during the war in nearby Brussels closely, but what exactly had happened there
that would have made them apprehensive about Einstein, Carl or Albert? What role did Carl
Einstein play in the communist movement? Were there other occasions in which the two
Einsteins were confused?  By answering these questions, we will gain new insight into the
wider reception of Albert Einstein and his theory of relativity.

Carl Einstein (figure 7) was born in Neuwied, near Koblenz, in 1885. He grew up in
Karlsruhe, where his father became the principal of a school. His parents were Reform Jews;
his family was unrelated to Albert Einstein's.[81] Carl, like Albert, had dropped out of high
school (*Gymnasium*).  In 1904 he went to Berlin to study; he enrolled in courses in
philosophy, art history, history, and classical languages.  As a young man, he shared with
Albert a fascination for Ernst Mach's ideas; Carl was particularly impressed by Mach's
notion of an "element of sensation"--he later would argue that the Cubists had captured
particularly well a Mach-like sense of observation, just as he himself would try to capture this
in his own literary writings.[82]  He ended his university studies in 1908 without graduating (so
he actually was not a "Dr."), and set out to make a living for himself as an author and critic.
His experimental novel, *Bebuquin oder die Dilettanten des Wunders* (*Bebuquin or the
Dilettantes of Wonder*), which appeared in various installments beginning in 1908 and as a book in 1912,[83]quickly made him distinctly noted in the literary world. In 1913 he married Maria Ramm; the marriage would be dissolved in the early 1920s.

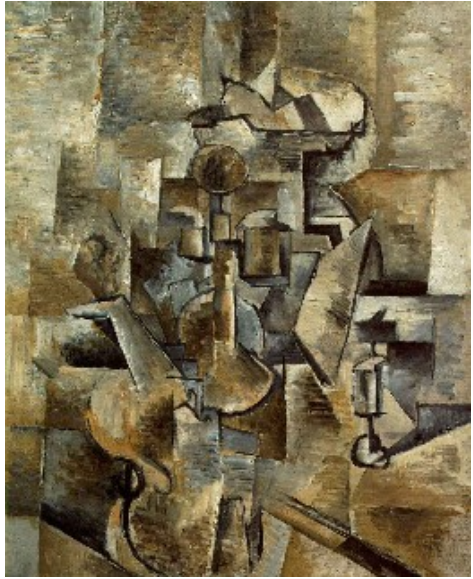

Figure 8: Georges Braque, *Violin and candlestick*, 1910. Source: Wikimedia.

Carl Einstein began his career as an expressionist: He belonged to the group of authors and artists who wished to revolt against the bourgeois values and culture of Wilhelminian Germany, and did so by presenting new, antinaturalist forms. [84] Art had to resist the unfree aristocratic world and the state's rationalizing forces: Everyone should aspire to a free and communal society, and many expressionists, himself included, would feel at home with anarchist views. He considered himself to be a radical revolutionary, but before 1915 his desired revolution was primarily cultural and ethical, rather than concretely political.

According to Carl Einstein, most post-Impressionist painting had become too decorative and aspired too little to an objective ideal. To arrive at true art and a true philosophy, he demanded that the usual way of representation be destroyed, and that a new representation be formed from the perspective of idealistic aesthetic interpretations. He would demand a similar interaction--a continual tearing down and building up, a permanent struggle through crisis and dialectic criticism--for his own writing and for his own life, even for his daily life; Carl Einstein was engaging, provocative, full of unrest, and characterized by a jerky intensity.[85] He had the personality of a revolutionary, and wished to produce and promote revolutionary art. After the publication of *Bebuquin*, he wanted to write in a still "more radical" way "to create a completely new mind for this age." There could be no compromise: "I have to be unbearably fanatic, almost impossible to put up with. Otherwise I consider my work a failure."[86]

Carl Einstein was an early and strong supporter of Cubist art (figure 8). In his opinion, it was the perfect starting point from which painting could move beyond the decorative: Cubism opposed the stylized and scientific "forgery" that was the norm, and instead managed to capture experiential reality with its suggestions of spatial and temporal extension.[87] He would regularly travel to Paris and there met Pablo Picasso, Georges Braque, and Juan Gris. They, and the circles to which they belonged, had begun to appreciate sub-Saharan African art, even though European art criticism until then had largely dismissed African sculpture as primitive. Carl Einstein would be one of the earliest critics to discuss it in the context of Western aesthetic norms in his book, *Negerplastik*,[88] which at the same time outlined a wider programmatic view for Cubism. Its publication in 1915 established his reputation as an art critic.

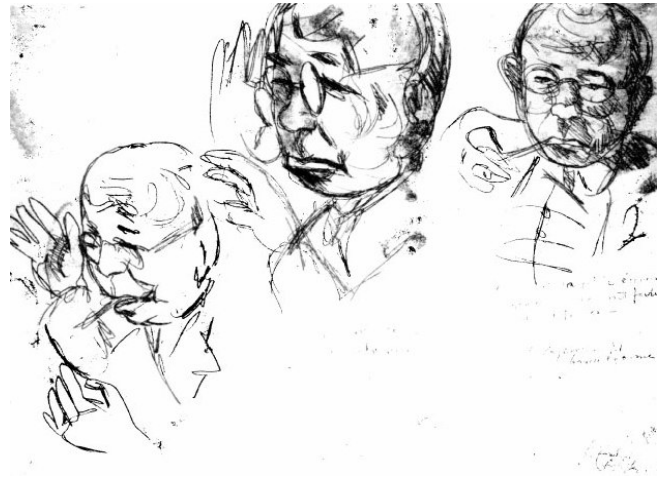

Figure 9: Carl Einstein as sketched by Rudolf Grossmann (1882-1941) in 1921. Below the middle image, Carl Einstein had written: "Et quand même c'est moi qui a fait la révolution en Belgique!" Source: Carl Einstein, *Werke* (ref. 118), inside front cover.

*Negerplastik* came out while Carl Einstein was serving in the German army. He had enlisted as a volunteer, inspired by the euphoric "Spirit of 1914" that had swept through Germany at the outbreak of the Great War. His enthusiasm was not exceptional among expressionists: the war was welcomed by many of them as the much-desired sharp break with an odious and complacent peace.[89] After serving in the Alsace region, he was ordered to Brussels in 1916, the center of the German administration on the Western front. He worked there in a subordinate position in the colonial department of the German civil administration. Soon his revolutionary spirit turned, however, to concrete political action:[90] He became one of the prominent figures in the Soldiers' Council of Brussels, the *Soviet* of German soldiers who revolted against their former commanders and ruled the city for a short and turbulent period in November 1918.

The Soldiers' Council marked the end of the German occupation of Belgium. By August 1918, after an often brutal occupation, it had become clear that Germany stood in a lost position; it began to withdraw its troops in mid-October. The troops then turned against their commanding officers, as in Germany itself, and on November 9, 1918, the day on which the abdication of Kaiser Wilhelm II was announced, the Soldiers' Council took over the civil administration in Brussels.[91] For a few days, the Council attempted to be the city's central authority and directed the withdrawal of troops; Carl Einstein (figure 9) was its press officer. He also would function as its representative in consultations with Belgian officials and emissaries from neutral countries.[92] He thus was one of the key figures in the operation of the Soldiers' Council.

The Council's revolutionary character was exhibited in the way in which Carl Einstein dealt with his former superiors in the deliberations on the withdrawal of German troops. On the first day of its operation, neutral diplomatic representatives of the Netherlands and Spain, the president of the Belgian national relief committee, and the former leaders of the German civil government--the latter were now stripped of their uniforms and regalia--were assembled for a meeting with the Soldiers' Council. Local city officials who were present recalled the unusual event:

> After a little while a door swung open, and someone in civilian clothes, wearing a monocle and with half of his skull wrapped in bandages, walked up. This strange fellow was called Einstein; he was art critic in Paris and, just until yesterday, he was a small-time junior employee in the German administration of Wallonia in Namur, under the command of Governor Haniel…. "I am," he said [without looking at the German officers], "a member of the Council of Soldiers and Workers...." "Von der Lancken [chief of the political department of the German administration, who Einstein indicated by pointing his finger at him, while omitting his title of baron] told me that I could find everyone here."

After this preamble, Einstein straightened himself. While walking around, looking angrily at Moltke, Rantzau and the others [members of the local German command], he said in a loud voice: "This abysmal imperial regime has come to an end. As of today, the Council of Soldiers and Workers intends to succeed the oppression and tyranny that has weighed on Belgium for so long with a humane and loyal regime. The German soldiers don't want anything except to leave Belgium as quickly as possible. Let's allow them to leave and try, from all sides, to avoid any conflicts! I have been commissioned by my comrades of the Council to see if there is any measure that we can take in the interest of resupplying Belgium during the period of the evacuation."[93]

At this, the Spanish envoy indicated that the Belgian national relief committee had organized the provisioning of the country, and that it would be most helpful if its workers would not be hindered in any way. Einstein turned to Count von Moltke:

"Off to the telephone! Call up the general in charge of waterways and tell him that I give orders that things should happen as has just been said!" He added, with his gaze fixed on his compatriots: "The Council of Soldiers and Workers at present calls the shots, and that implies that its orders are to be followed up on the spot! The odious oppression that this population has had to suffer has to stop. All Belgian civilians who have been arrested and jailed during the occupation have been released on my order."[94]

The relief committee's leader concurred with the suggestion of the Spanish emissary, but also complained that some of his supplies had been requisitioned by the occupying forces. Einstein reacted furiously:

Einstein, in a menacing manner, turned to the Governor of Wallonia, Haniel, to whom he had been quite subordinate only yesterday. Shaking his fist, he yelled at him: "Sir, I am very surprised by what the president of the National Committee has just said. Why are you not acting in line with my orders? Take note of those I am giving you right now!" While the gang of barons and counts listened without uttering a word, ex-Governor Haniel humbly took a notebook from his pocket and meekly wrote Einstein's directions down. [95]

Three days later, on November 13, Carl Einstein was still excited about the latest developments, in particular about rumors that the troops at the front were fraternizing. Later that day, however, he seemed worn out and dejected, and advised a friend to leave Belgium as quickly as possible.[96] The Council's rule soon collapsed.

Sybille Penkert, in her most useful, early study on Carl Einstein, has suggested that he and the Council played a constructive role toward a more or less orderly evacuation of Brussels.[97] Yet, some who had direct dealings with Einstein have painted a different picture: The Dutch Ambassador, M.W.R. van Vollenhoven (yes, another namesake), for example, wrote in his memoirs that Carl Einstein, "not knowing anymore how to get out of trouble," at one point had asked him and the Spanish ambassador to mediate between the Soldiers' Council and the Belgian authorities--thus, in effect, disqualifying Einstein's own role. Various factions in the German army had begun fighting each other, and the Soldiers' Council kept issuing conflicting orders, while the German troops fired on the Belgian population; Einstein, as an excuse to van Vollenhoven for their assault, suggested that the Belgian citizens had turned on the German soldiers. By now, however, the Dutch ambassador had seen enough of the chaos, and he promptly sought a transfer of power from the Germans to local Belgian authorities, as indeed was effected on November 13. One of the Belgians' first measures was to rearrest some of the criminals whom Carl Einstein had released.[98]

Thus, the events in Brussels that were contained in the Dutch intelligence report are now clarified. But what about "the Red Countess Olga von Hagen"? Her actual name was Agathe von Hagen (*ca.* 1872-1949), and she was indeed a countess. She and Carl Einstein had met through Gottfried Benn, the expressionist author and close acquaintance of Einstein who also was in Brussels. Von Hagen was employed as a social worker by the German authorities and became Einstein's partner for more than a decade; an acquaintance called her "sympathetic and humane," and Benn found her "royal" and "wonderful."[99] She was well connected: when Carl was apprehended by the police in Bamberg, he gave her name, along with that of the National Minister of Internal Affairs (*Reichsminister des Innern*) Eduard David, as references for identification purposes; David was a close friend of von Hagen.[100] She and Carl Einstein made an odd couple; they have been characterized as the personification of the contradictions between charitableness and criticism. Whereas the excitable Einstein could be "uncouth," Countess von Hagen appeared "calm," even during the German collapse and revolutionary storm that swept through Brussels.[101]

We can safely conclude that Dutch officials in The Hague had confused Albert Einstein with Carl Einstein. The latter would have been in their files, since in nearby Brussels he had evolved into a communist revolutionary leader. The rebellion that led to the Soldiers' Council in Brussels should be seen as an integral part of the German Revolution of November 1918. Its leaders were also strongly inspired by the Russian Revolution of the preceding year; they had adopted both its vocabulary and its Soviet ideology. Carl Einstein left Brussels both with the conviction that political action inspired by that ideology could be successful, and with the will to put this conviction into practice. After the collapse of the Soldiers' Council in Brussels, he quickly departed for Berlin in the middle of November,[102] filled with an urgent desire for revolutionary political action.

**Dutch Revolutionary Fears and Professorial Appointments**

On Armistice Day, November 11, 1918, Carl Einstein had opened the "official" meeting of the Central Soldiers' Council in Brussels.[103] That same day, only 140 kilometers to the north in Rotterdam, the leader of the Dutch Social Democratic Labor Party (SDAP), Pieter Jelles Troelstra (1860-1930, figure 10), had proclaimed at a party meeting that "the working class in the Netherlands at this moment seizes political power. It will have to constitute itself as a revolutionary power," and he had repeated his proclamation the next day in parliament.[104] Despite the Dutch government's best efforts, the revolutionary fever had not been contained to Brussels or to Berlin, but also had arrived in the Netherlands. There were substantial differences, however, and by the end of the week it was clear that a Dutch revolution had failed; in fact, it had never really got off the ground.

The majority of the SDAP and union leadership had not been in favor of Troelstra's move, and did not follow suit. The country had held its first general election just a few months earlier, and only twenty-five percent of the vote had gone to the socialists; in these circumstances, a revolution could hardly expect broad popular support. Troelsta, however, closely following events in Germany, was afraid that the SDAP would lose the revolutionary momentum to communists and anarchists. His proclamations created quite an upheaval, but he soon had to admit that he had been "mistaken." The SDAP subsequently passed a motion stating that it would remain a legal and democratic party, and would only join any revolution if the majority of the Dutch working class supported it; this, however, was unlikely to transpire because of its large membership in confessional groups. Unions associated with these groups also had rallied to resist the revolution, and what became known as "Troelstra's mistake" soured political relations between socialists and other political parties for a long time.

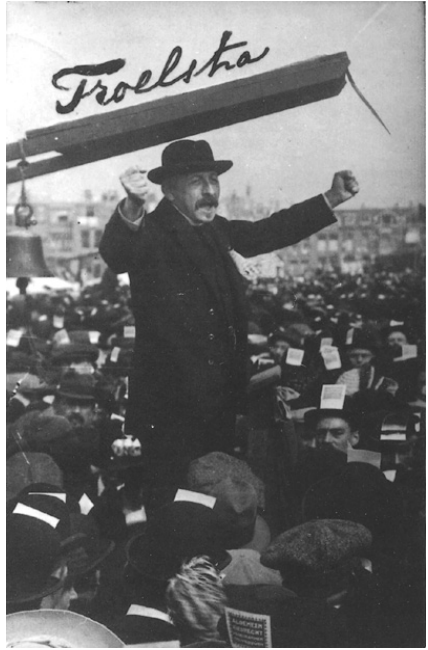

Figure 10: Pieter Jelles Troelstra (1860-1930), Dutch socialist leader. Source: Wikimedia.

Still, regardless of Troelstra's failure, the Dutch Cabinet, in which Minister of Education de Visser served, had been jolted by it, and had become deadly afraid of revolutionary councils and their leaders, such as Carl Einstein:  Reports that Russian Bolsheviks had allocated extensive reserves for bankrolling "propaganda" services in Western Europe further fueled anxieties, and some Dutch Cabinet members feared a renewed effort by Troelstra in the near future.[105]

Many of the Dutch government's social policies were directed at heading off such revolutionary attempts. Christian politics usually worked toward this end by co-optation: By promoting the organization of labor movements under a Christian banner, and by piecemeal improving the social plight of the working classes through legislation, one tried to keep the working poor inside the various churches and their affiliated organizations. Minister de Visser himself, earlier in his career when still a clergyman, had formed a local chapter of a Christian labor organization that aspired to gradual social change; his belief had been that only the Christian faith held the promise of harmonious coexistence of the proprietary and working classes.[106] Obviously, the leaders of the Christian people's parties, including de Visser, thus very much would wish to keep revolutionary agitators out of professorial chairs.

Albert Einstein's difficult appointment procedure was not the only example. In April 1920, Ehrenfest informed Einstein of his hunches as to why Einstein's appointment was being held up:

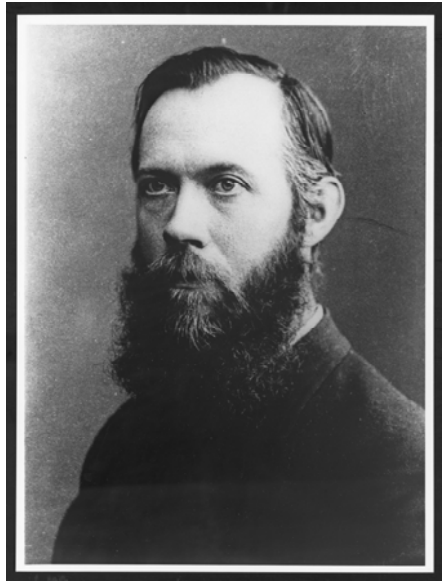

Figure 11: Astronomer and socialist Anton Pannekoek (1873-1960), around 1906-1910.
Source: University Library, University of Amsterdam.

*Probably*, the Minister [de Visser], owing to the Bolshevik actions (as leaders and propagandists) of some professors here in this country, is having confirmed in Berlin that you are not another bolshevism-propagandist. The more famous you are, the more care he needs to take that you are not possibly some agitator, before pressing you to his breast![107]

A particular case that Ehrenfest would have had in mind would have been the failed appointment in Leiden of the astronomer Antonie Pannekoek (1873-1960, figure 11), which fell through in the spring of 1919, less than a year before Einstein's appointment came up. The case of Pannekoek is revealing, as historian David Baneke has shown,[108] in particular owing to the heightened sensitivities as a consequence of Troelstra's actions.

Pannekoek was both a prominent astronomer and a prominent theorist of socialism, and he insisted that the two parts of his intellectual persona were strictly separate. In astronomy he was known, for example, for his work on the detailed structure of the Milky Way; as a socialist, he stood to the left of Troelstra's SDAP: In his view, the revolution had to arise directly out of the people, and he opposed strict party discipline. His contributions to socialist theory were greatly appreciated: In 1906 the German Social Democratic Party had invited him to become a lecturer at its party school in Berlin, and Lenin, for example, had been an admirer of his work.[109] Pannekoek would work as an ideologue in Germany until 1914 when, at the outbreak of the Great War, he returned to Holland and took up a teaching post in a secondary school, along with a position as an unpaid lecturer (*Privaatdocent*) at the Leiden Observatory.

On March 3, 1918, Ernst Frederik van de Sande Bakhuysen, the elderly director of the Leiden Observatory, died. Local administrators proposed Willem de Sitter as his successor, who insisted that two adjuncts also should be appointed: the Dane Ejnar Hertzsprung (1873-1967), who would head the Observatory's astrophysical department, and Pannekoek, who would head its Meridian Department. In October 1918, Minister de Visser accepted de Sitter's proposal--just a few weeks before "Troelstra's mistake"--but delayed its execution until further notice. Then, early in 1919, articles began to appear in the Dutch press on Pannekoek's communist connections: some alleged that he had received funds from Russia for propaganda purposes in Holland, and that he would have been elected as honorary president of a revolutionary Soviet Council in Hungary, where Béla Kun's revolution was underway.

There was no basis of truth to the first story, but there was some substance to the second--even if Pannekoek had had no hand in his honorary Hungarian election. De Sitter lost his patience: "What are you really," he asked Pannekoek, "a communist or an astronomer"? Pannekoek explained that he had not been politically active since his return to the Netherlands. He had completely "returned to science"; it was not his fault if the Dutch government wished to sacrifice science to political interests.[110] The appearance of another newspaper report, pointing out that a known communist was about to be appointed to a prestigious post in the Leiden Oberservatory, sealed his fate: The next day, on May 3, 1919, de Visser announced that "under no conditions" would he appoint Pannekoek, "in view of the national interest."[111]

Meanwhile, de Visser had accepted de Sitter. Leiden's curators subsequently dropped their lobby on Pannekoek's behalf; de Visser's refusal in The Hague had been too categorical, and could not be expected to be overturned, in particular in light of Troelstra's failed revolution. The Chairman of the Curators, de Gijselaar, summed up the sentiment regarding Pannekoek: "those red gentlemen never keep their word," since at the slightest sign of unrest "they become unhinged."[112] Socialists in parliament raised the turn of events with de Visser: They charged that he was conducting a "political inquisition in higher education," and besides, Pannekoek could hardly be expected to "raise havoc with the stars."[113] De Visser would not budge: this candidate supported the overthrow of the State, and thus was patently unfit as an educator--certainly at times when "every foundation is shaking."[114] A motion of disapproval of de Visser's actions did not get a majority. Pannekoek soon became lecturer at the University of Amsterdam, which as a city institution did not need de Visser's approval to appoint its faculty; furthermore, Amsterdam's city council had a solid contingent of communist and socialist members.[e]

In the case of Pannekoek's proposed appointment at Leiden, it appears that all parties involved were more or less informed about his communist background from the outset. It thus may seem surprising that de Visser initially agreed to it. However, this of course was a chair in the uncontroversial subject of astronomy, and Pannekoek had retired from publishing on political subjects. In any case, Pannekoek himself had feared all along that approval of his appointment would fall through in the end,[115] and indeed, in the wake of the Troelstra affair, the Dutch government threw out his candidacy when newspapers began to report on it. This course of events strongly suggests that the Dutch government may have feared adverse public opinion at least as much as any actual revolutionary threat that Pannekoek may have presented. In either case, after Troelstra's actions both the Dutch government and the public at large would not and did not accept the appointment of a communist at Leiden. Thus, given their fresh experience with Pannekoek, it is not surprising that Dutch authorities subjected Albert Einstein to such an elaborate and meticulous appointment procedure, and that he ran into similar problems, particularly since they initially believed that they were dealing with a far more dangerous person than a retired theorist of socialism. Only after it had been spelled out and confirmed again and again that this Einstein was a different sort of revolutionary than the other Einstein, could the former Einstein be appointed to a special chair in Leiden.

---

[e] The willingness of Amsterdam's City Council to appoint "red" professors was already in evidence when it made Gerrit Manoury (1867-1956) Professor of Mathematics at the University of Amsterdam in 1918. In 1920, de Visser requested that the University of Amsterdam's curators investigate the recent political activity of Manoury and Pannekoek (rumors were circulating that they had attended a meeting of the Comintern). The University's curators replied that they saw no need for such a step: Neither one had misbehaved or neglected their duties, nor had they been engaged in propaganda activities; see Knegtmans, "Politiek aan de Nederlandse universiteiten" (ref. 41), p. 114; for the City Council's stewardship of the Amsterdam Physics Department in this period, see A.J.P. Maas, *Atomisme en individualisme. De Amsterdamse natuurkunde tussen 1877 en 1940* (Hilversum: Verloren, 2001), chapters 3-4, pp. 89-158.

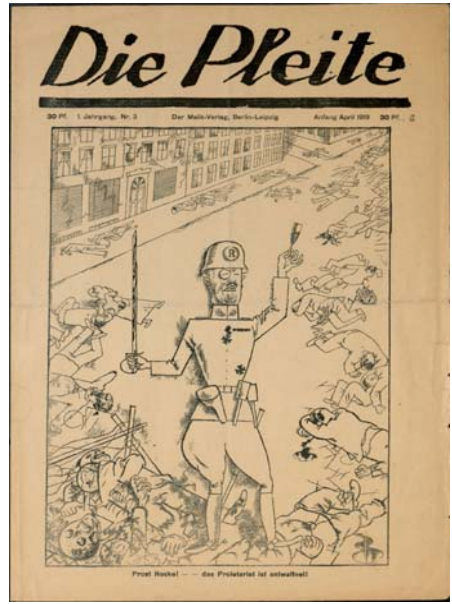

Figure 12: George Grosz (1893-1959), cover of *Die Pleite* (April 1919). The caption reads: "Cheers Noske—the proletariat has been disarmed!" Gustav Noske (1868-1946), Social Democratic Minister of Defense, was ultimately responsible for the violent quelling of the communist revolts in early 1919. *Die Pleite* was a leftist magazine to which both Grosz and Carl Einstein contributed.

**The "Communist Einstein"**

At the end of the Great War, while Carl Einstein hastily departed from Belgium, Albert Einstein's name was circulating in the Berlin press in relation to various democratic initiatives in Germany. On November 13, 1918, he gave a lecture at a public meeting that was attended by more than a thousand people and was prominently reported on. He spoke out in support of the November Revolution, but he strongly warned against a violent and undemocratic "tyranny of the Left."[116] Albert Einstein thus stepped forward as a prominent voice in the political arena, but not as an anarchist revolutionary.

Did only Dutch officials confuse Albert and Carl Einstein? Or were there other times and places where they could be or in fact were mixed up? Albert thought so: in an interview in December 1919 for the *Neues Wiener Journal*, he stated that, "In various newspapers I am portrayed as an emphatic Communist and anarchist, obviously due to confusion with someone who has a similar name. Nothing is farther from my mind than anarchist ideas."[117] As we will see, Carl Einstein's continued and visible role in a number of revolutionary efforts would produce ample opportunities to tangle up the two.

On his return to Berlin in the middle of November 1918, Carl Einstein immediately entered the revolutionary fray. He published communist appeals, for example, "To the intellectuals--One thing remains to be done: to make a communist society a reality,"[118] and he was involved in the violence that took place in the Berlin newspaper quarter (*Zeitungs-*

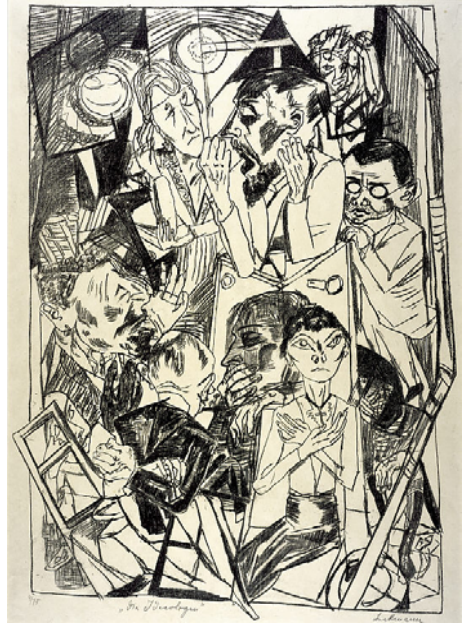

Figure 13: Max Beckmann (1884-1950), *Die Ideologen* (1919). Litograph number six in a portfolio of eleven called *The Hell*, documenting Beckmann's Dantesque journey through postwar Berlin. Heinrich Mann (1871-1950) is lecturing at this leftist salon while below him Beckmann, hand over mouth, closes his eyes in disgusted despondency. Carl Einstein stands behind Mann, stocially, and Agathe von Hagen (ca. 1872-1949) gazes upward, crossing her arms. Source: National Galleries of Scotland.

*viertel*) during the Spartacist uprising.[119] At its conclusion, on January 16, 1919, a day after communist leaders Rosa Luxemburg and Karl Liebknecht were murdered, the leftist paper *Freiheit* prominently reported on its front page the arrest of "the author Carl Einstein," and claimed that it had taken place without any probable cause.[120] Later, Carl was part of Berlin's brutal March battles (*Märzkämpfe*) that followed a general strike and involved summary executions by government-backed paramilitaries; afterwards he had to live as a fugitive, "fleeing from house to house," in the words of artist George Grosz (figure 12).[121]

Carl Einstein's name surfaced again in the press in April and May of 1919 as he lectured at the second Congress of Soviets in Berlin (figure 13). In the following month, just before he traveled to Bavaria, newspapers reported that Rosa Luxemburg's corpse had been recovered from a Berlin canal. Carl was one of six speakers at the ceremony commemorating her passing--a ceremony at which thousands were present. According to one account in the daily *8-Uhr-Abendblatt*, many Spartacists had spoken in a inflamatory manner at the event. The paper particularly singled out "the communist Einstein" (note: identified without his first name), who would have tried to incite the crowd to "pick up by surprise and kill," not only those responsible for Luxemburg's murder, but also those who had silently condoned it.[122] Denials soon appeared in *Republik* and *Freiheit* from attendees, and from Carl Einstein himself: The *8-Uhr-Abendblatt* had given an entirely false representation of what had transpired; the faulty report had been due, according to Einstein, to an attempt to slander his name and that of other communists.[123]

The events at the commemoration of Luxemburg resurfaced in a Munich newspaper that recounted the police apprehension of the "communist Einstein" (note: identified again without his first name) near Bamberg. Carl Einstein had been on his way to Nürnberg to lecture at a political rally, but on June 18, 1919, after his arrest and after the local police received confirmation of his identity, he was directed to return to Berlin.[124] The Berlin press, in their accounts of his arrest, added to the story new twists ("Einstein would have loaned his passport to Levien") and turns (there would have been a "campaign against Einstein"). Most surprising was a new claim that Carl Einstein's supposed incitement at Luxemburg's graveside had been fabricated by press officers of a paramilitary *Freikorps* that, in fact, was under the command of those responsible for her murder.[125]

Relevant for our purposes is that with the repeated mention of a Berlin communist Einstein in the press (at times with the omission of a first name), it could be expected that Albert Einstein would indeed appear to some as a leftist radical, or could be completely confused with Carl Einstein, as in The Hague. His unfounded reputation as a communist, which to the chagrin of his wife accompanied Albert Einstein in the early years of the Weimar Republic[f] seems attributable, at least in part, to Carl Einstein's public presence.

After the most turbulent months of the Revolution of November 1918 had passed, Carl Einstein remained a prominent communist activist; at times, however, he would suspend his political agenda or revise his positions. He quickly became disappointed in developments in Germany, but persisted in a personal and artistic rebellion. His alter ego, Bebuquin, expressed in a sequel his frustrations with the outcome of events in 1919:

> [The] atmosphere of the Berlin revolution … consists of a lack of leadership; careerism of socialists; salonbolsheviks; aristos etc. against the university, against the Nazis with their paid professional revolutionaries, faced on the other side by communist functionaries.[126]

Carl Einstein's disenchantment with Weimar politics strengthened his renewed immersion in art;[127] his iconoclasm in the arts, of course, in turn also remained a way to continue his resistance to bourgeois values.

In July 1921, Carl's work, *Die schlimme Botschaft* (*The Bad News*),[128] was published. In it Jesus Christ was crucified again, but this time in the acerbic atmosphere of postwar Germany. The work initially appeared to fall flat, but it eventually drew the attention of rightist circles, which led to a spectacular blasphemy trial. Carl Einstein was convicted, and, by now very publically visible, he was attacked as a Jew and socialist.[129] This event marked the end of his literary career. He continued as an ever-more-prominent art critic, however, and returned to Paris in 1928. He did not terminate his political engagement, even if his involvement with elitist art seemed to contradict his role as an intellectual who stood on the extreme left. In 1936 he joined the Spanish Civil War and fought on the side of the anarcho-syndicalists against Franco. After their defeat, he returned to France, where at the outbreak of the Second World War he was placed, as a German, in an internment camp. Upon his release during the German invasion, Carl Einstein (figure 14) fled to the Pyrenees where in the end, in 1940, he took his own life.

## Mirror Images

We have seen a number of possible and actual occasions in which Albert Einstein and Carl Einstein were misidentified. This has shown, among other things, why Albert Einstein's Leiden chair was so much delayed, and how it could be that many saw him as politically much more radically leftist than he actually was. This public misconception likely would have been reinforced by the circumstance that both Einsteins were perceived as modernist intellectuals.[130] In many ways, Albert and Carl were not just namesakes, but also each other's mirror image. Albert, too, was labeled a "revolutionary," even if he himself may not have chosen that title.[131] His revolution was in science, not in politics or art, and it succeeded, while many of Carl's attempts, who unlike Albert strenuously sought after the revolutionary epithet, failed. By 1918 Albert played the role of revolutionary reluctantly--yet his revolution was eventually as much in the public eye as those Carl pursued. This suggests that we could learn more by a further comparison of the two, which I now will undertake. The emphasis will be on what it can teach us about the reception of relativity.

---

[f] Albert Einstein's "red" reputation would stay with him, even after his permanent departure for the United States in 1933; see Fred Jerome, *The Einstein File: J. Edgar Hoover's Secret War Against the World's Most Famous Scientist* (New York: St. Martin's Griffin, 2002).

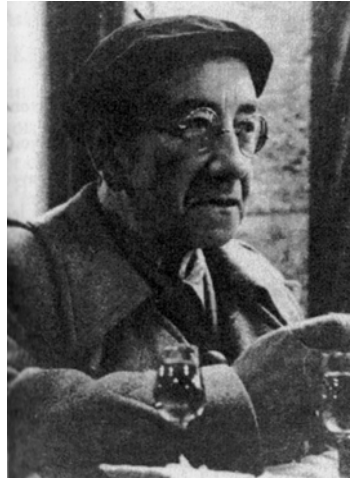

Figure 14: Carl Einstein in 1939. Source: Bibliothek der Freien, Berlin.

Let us first return to Paul Weyland and his very vocal opposition to relativity in the summer of 1920. The metaphor of mirror images suggests itself in his case directly--so much so that one wonders whether Weyland in fact *did* confuse Albert and Carl, or, rather, chose to play on their possible confusion. For example, a year earlier, in May 1919, Carl lectured at the Berlin Philharmonic on the "political responsibility of the intellectual" in a series organized by the Soviet newspaper, *Räte Zeitung*, that was to mark the founding of a Soviet Society (*Räte Bund*); the Society's goal was to promote and deepen Soviet ideology.[132] Carl was one of two speakers that night. These coincidental similarities with Weyland's event may not actually have been coincidental: Weyland's later Philharmonic event and his Working Society look like a mirror image of Carl's event, even if political gatherings at the Berlin Philharmonic were not uncommon. Perhaps Weyland chose the location and format of his evening on purpose, as tacit reference to Carl's appearance? We also saw that Weyland drew explicit parallels between an alleged marketing of relativity and Dadaists' practices. Did Carl Einstein also have any Dadaist credentials?

Carl Einstein is not usually identified as a member of the Dada movement: He did not participate in Dada events, nor did he publish in Dada periodicals. Nonetheless, during 1919 and 1920 he collaborated intensively with George Grosz, and the brothers Wieland and Helmut Herzfeld (the latter used the pen name John Heartfield), all of whom were prominent members of the Berlin Dada group. Carl Einstein shared his desire for political revolution and antipathy to bourgeois culture with them, even if they differed on what political and social role art ought to play.[133] He contributed articles to the Herzfelds' politically-oriented periodical *Die Pleite* (see figure 12 above), and together with Grosz he composed the satirical magazine *Der blutige Ernst* (see figure 15 below). Grosz and Carl Einstein declared that their magazine "pinpoints Europe's sicknesses, catalogs the total collapse of the continent, combats the deadly ideologies and institutions that caused the war, and confirms the bankruptcy of Western culture"--in its pages the tone and themes of George Grosz's artwork and Carl Einstein's writing complemented each other naturally.[134] Even if *Der blutige Ernst* was not a Dada publication, Grosz was already well known as one of Dada's leading Berlin representatives; some of that reputation would easily have rubbed off on Carl Einstein.

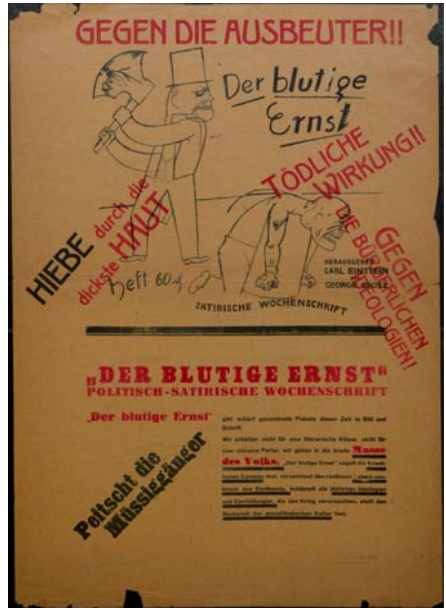

Figure 15: George Grosz and Carl Einstein, advertizing pamphlet for *Der blutige Ernst* (1919).

*Der blutige Ernst* ceased publication in February 1920, and Carl Einstein, despite his collaboration, appears not to have warmed much to Dadaist art and its typical photomontage technique. The Dada group, in turn, seems to have had a lukewarm opinion of Carl Einstein's views on art. [135] In a photomontage entitled *Pablo Picasso: La vie heureuse. Dr. Karl Einstein gewidmet* (*Pablo Picasso: The Happy Life. Dedicated to Dr. Karl Einstein*, figure 16), which was included in the first International Dada Fair, Grosz and Heartfield pasted together a Picasso reproduction and the eyes of a bourgeois resembling Carl Einstein, along with a *Freikorps* officer and the word "Noske," the name of the reviled social democrat who was held accountable for much of the government's violence of the preceding year. The composition rendered Carl Einstein's Cubist fixations as an expression of odious establishment tastes.

The Dada fair generated wide media attention, as intended. Rightist dailies were outraged by its artworks, and the exhibition was subjected to a lawsuit for defamation of the German army.[136] It was held in the center of Berlin, close to the Philharmonic, and took place from June 30 to August 25, 1920. Weyland thus had scheduled his anti-relativity event exactly on the eve of the end of the Dada fair. Carl was not the only Einstein depicted at the fair; also on display was Hannah Höch's collage, *Schnitt mit dem Küchenmesser. Dada durch die letzte weimarer Bierbauchkulturepoche Deutschlands* (*Cut with a Kitchen Knife. Dada through the Last Weimar Beer-Belly Cultural Epoch of Germany*, figure 17). In it a front-page photograph of Albert Einstein from the *Berliner Illustrirter Zeitung* figured prominently, together with pictures of, among others, Kaiser Wilhelm II, rightist Putschist Wolfgang Kapp, artist Käthe Kollwitz, and Karl Marx. Those whom Höch viewed favorably--those who opposed conservative values and reactionary politics--were placed close to the word "Dada," as was Albert Einstein (figure 18), despite his own somewhat traditional tastes in art;[137] those whom Höch disliked, like the Kaiser, were placed next to the word "anti-

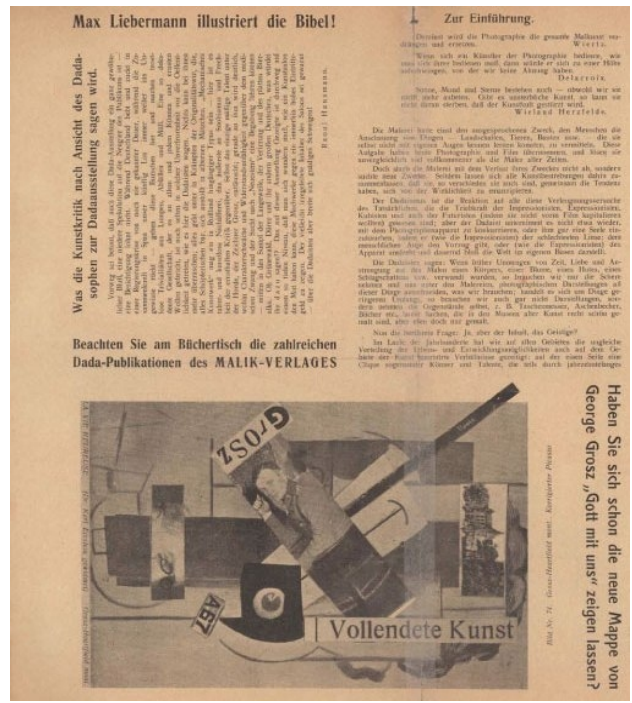

Figure 16: A reproduction of George Grosz and John Heartfield's *Pablo Picasso: La vie heureuse (dr. Karl Einstein gewidmet)*, 1920, was contained in the *Erste internationale Dada-Messe: Katalog* (Berlin: Kunsthandlung Dr. Otto Burchard, 1920), p. 2. The original of the photomontage is believed to be lost, the only known image of it being this catalogue image.

Dada." The artwork represented current artistic and political revolutions; it particularly drew attention to women's right to vote, and how this would alter Germany's "beer-belly culture." Almost entirely composed of newspaper clippings, it also captured the hyped atmosphere of Berlin's daily press, and its images of high-rise buildings and factory machinery depicted the modernist spirit that many conservatives found deeply objectionable.[138]

The names and images of both Einsteins thus circulated at the Dada fair, and they appear as an entangled reflection of each other: as bourgeois and antibourgeois, and revolutionary vanquisher and sellout. In the end, whether Weyland actually had jumbled up Albert and Carl Einstein is not of immediate relevance; their dual roles, given the uproar that both relativity and Dada were generating, serve to illustrate a larger context. Deliberately or not, Weyland saw an interest in portraying Albert as a kind of Carl Einstein; he chose to smear Albert Einstein with the scandalous reputation of Dadaist art. The tactic worked: Max von Laue, present at the Philharmonic event, summed up Weyland's rhetoric in a letter to Arnold Sommerfeld as follows: "Einstein would be a plagiarist, supporters of relativity publicity makers, and the theory itself Dadaism (that word was really put forward!)"[139]

Weyland thus not only played on anti-Semitic sentiments in his demagogic rant, he also attempted to capitalize on the negative reaction that the revolutions in art and politics precipitated--revolutions of which Carl Einstein, speaker at Rosa Luxemburg's graveside and coeditor with George Grosz, formed the vanguard. Weyland wished to mobilize these negative sentiments against another onslaught on cherished values, namely, the revolution that wished to topple established scientific knowledge, driven by yet another revolutionary Jew, Albert Einstein. Weyland's phantom Working Society thus was supposed to labor for the "Preservation of Pure Science." In turn, by co-opting the anti-relativist stance, he also intended to enlist support for his larger desires: to resist the Revolution of November 1918 and modernist culture. Weyland's whipping up of anti-relativistic sentiments thus was a natural complement to a broader politically and culturally reactionary agenda. All in all, the above suggests that part of the German resistance to relativity should be seen as resistance to the revolutionary spirit of the postwar years. For Weyland, at least, it was only a small step from anti-revolutionary to anti-relativist, and his opposition to relativity brought science into an unfamiliarly shrill anti-revolutionary discourse.

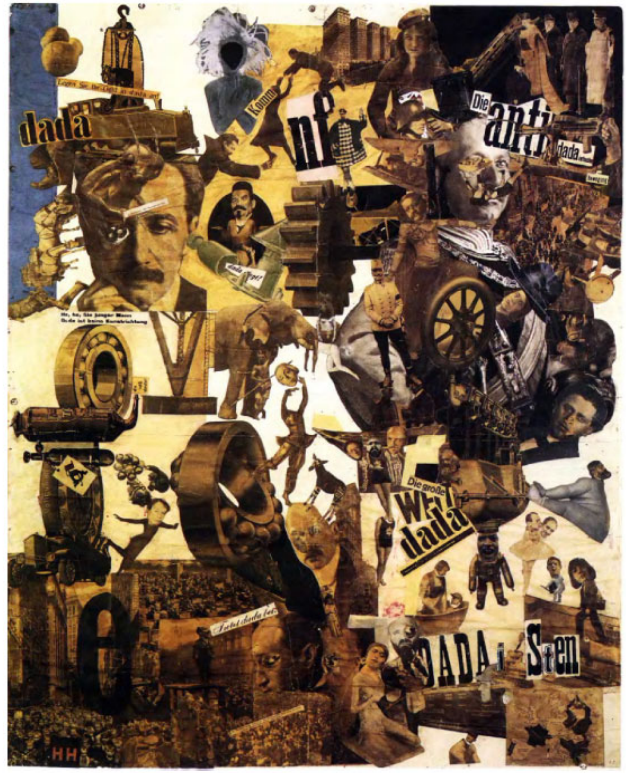

Figure 17: This collage by Hannah Höch, *Schnitt mit dem Kuchenmesser* (full title in text), was featured in the First International Dada Fair, held in art dealer Otto Burchard's gallery in downtown Berlin from June 30 to August 25, 1920. The collage featured not only Albert Einstein, but also many other public figures, including the first president of the Weimar Republic, Friedrich Ebert (1871-1925), the reactionary Putschist Wolfgang Kapp (1858-1922), Kaiser Wilhelm II (1859-1941), socialists, Dadaists, and Karl Marx (1818-1883). Source: Bildarchiv Preussischer Kulturbesitz, Berlin.

Weyland, of course, was far from the only opponent of relativity and "anti-relativists," as they are known collectively in the literature, can hardly be identified as a unified group with a single agenda. The identification applies to a rightist experimental physicist like Philipp Lenard, to the philosopher and converted Jew Oskar Kraus, and to amateur scholars like the engineers Rudolf Mewes and Hermann Fricke who wished to defend their own private solutions to the riddles of the universe. Milena Wazeck, in her recent insightful study of the phenomenon, has pointed out a particular feature that helps to explain the vehement character of some of the opposition that Einstein encountered: many, if not all of these opponents of Einstein fought against their marginalization by the specialization and professionalization pressures of established science. Relativity, as the most visible representative of these pressures, was an obvious target for dissent.[140] Wazeck's analysis thus differs from the earlier literature, which mostly pointed to the political dimension of the opposition to Einstein.[141]   However, this weakening of explicit socio-political factors also seems to reduce the possibility of understanding the actions of German anti-relativists in their Weimar context. It may be argued further that Wazeck's point of view too easily follows anti-relativists in their self-fashioning as *unpolitisch* or apolitical: Anti-relativists often wished to present themselves as scholars who were interested exclusively in matter-of-fact discussions on the merits of Einstein's theory, far from the political fray.  This self-qualification sounds dishonest, however, since it conflicts with the vehemence of their criticisms, and the political overtones of many of their statements.[142]

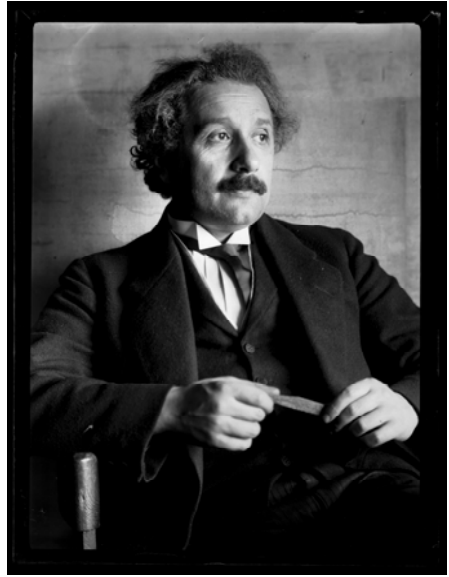

Figure 18: Albert Einstein in 1921, photograph by Ferdinand Schmutzer. Source: Bildarchiv, Austrian National Library.

In 1905 Einstein regarded his light–quantum hypothesis as revolutionary, not his theory of relativity,[143] which could be viewed as a reformulation of Lorentz's familiar electrodynamics. Nevertheless, relativity was soon widely seen as a revolutionary theory in the early history of its reception. This perception deepened, and spread throughout society at large after the extensive press coverage of the eclipse results of 1919 that confirmed Einstein's predictions.[144] Along with the theory's contentions about space and time, this perception can hardly seem surprising, given the particular historical moment in which the eclipse results exploded onto the world scene: their dramatic presentation came at a time when there were political revolutions in many European nations. At the same time, modernist perspectives were revolutionizing the arts.[145] In such an atmosphere, scholars as well as the public at large would be more prone to view Einstein's theory of relativity as "revolutionary," rather than as a reinterpretation of Lorentzian electrodynamics: in revolutionary times, one expects revolutions. Albert Einstein's personal background--Jew, democrat, and pacifist--would have aided that identification, particularly in Germany, whether or not he was being confused with Carl. From this perspective, it thus should not be surprising, for example, that the *Berliner Illustrirter Zeitung* chose to caption Einstein's image with the claim that his work "signifies a complete revolution of our understanding of Nature."[146]

When tabloids announce yet another revolution, anti-revolutionaries find a new cause. Anti-relativist Ernst Gehrcke (figure 19) complained about a "revolutionary wave in the sciences," and he expressed the hope that "also in the development of the sciences, the thought of evolution" would regain the momentum.[147] He believed that

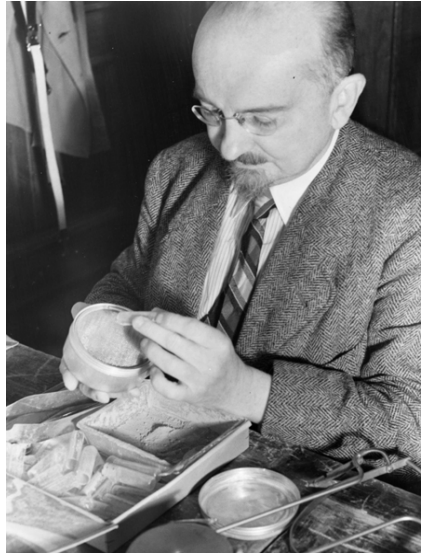

Figure 19: Ernst Gehrcke (1878-1960).

> [After] the end of the war, and in Germany the successful political overthrow of the State, the psychological moment for revolutions also in art and science seemed to have arrived, and first cautiously, then increasingly more forcefully, the full propaganda campaign for the theory of relativity was rolled out in the daily press [figure 20].[148]

The connotations of Gehrcke's words were not lost on his readers on the extreme right: The charge of propaganda was a clear code for them to pick up that relativity was to be spinned as yet another plot played out by yet another revolutionary Jew.[149] Although Gehrcke liked to present his position as apolitical, the words he chose and his analysis suggest otherwise--as does his repeated and unqualified identification of Albert Einstein as an Independent Socialist (as a supporter of the breakaway *Unabhängige Sozialdemokratische Partei Deutschlands*), that is, as a leftist radical.[150]

What was true for Weyland and Gehrcke was not true for all German anti-relativists: Wazeck is correct in identifying as a common denominator of the larger group's opposition the need to resist a perceived marginalization in current science. Yet, the above indicates that some of the most visible anti-relativists also feared marginalization of their social and political positions and conservative cultural values,[151] owing to modernist forces, the November Revolution, and the more radical pressures that loomed beyond. In this perspective, the angst and actions of the anti-relativists in the late 1910s and early 1920s were as much an expression of their resistance to threatening changes in broader social domains as to those in the sciences, or in physics proper; they served, in particular, as a reaction to the democratic left and modernist culture that seemed to dominate in the early years of the Weimar Republic.

Both Albert and Carl had warmly welcomed the collapse of Wilhelmine Germany. Albert saw his wishes come true with the creation of a parliamentary democracy, whereas Carl wished for and labored toward a true Soviet republic. Both Albert and Carl were equally offensive to those who wished to see neither come to pass; to obstruct Albert's theory most effectively anti-relativists such as Weyland and Gehrcke chose to portray him as much as possible as someone resembling Carl. In turn, the venomous opposition to relativity theory itself was due, at least in part, to the fear that "revolutionary" political ideals, communist or democratic, and modernist values, expressed most clearly in the arts, were gaining the ascendancy.

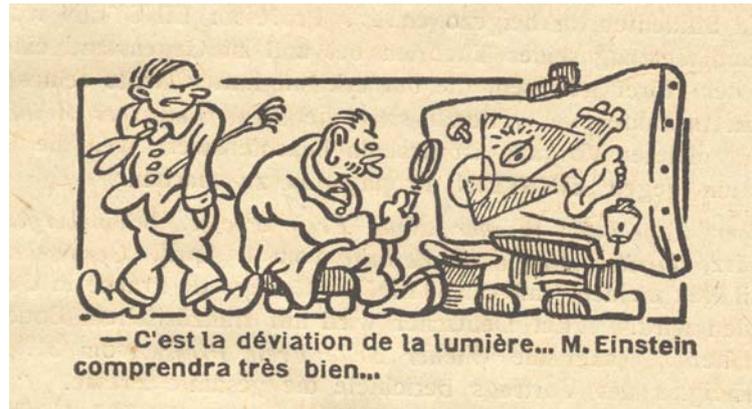

Figure 20: Cartoon from *Le Journal* (March 29, 1922). In the caption, a modernist artist says to a bewildered spectator, whose high hat is on the floor: "That is the bending of light… Mr. Einstein will understand very well…" Source: reproduced in Gehrcke, *Massensuggestion* (ref. 148), p. 77.

**Conclusion: Relativity and Revolution**

The history of Einstein's Leiden professorship is revealing in several respects. Appointing Einstein should not be seen as separate from the local political context of the Netherlands, just as it has suggested a new angle--the dual perspective of Albert and Carl Einstein--from which to reconsider the negative reception of relativity in some German circles; here, too, the political context was responsible for perceptions of Albert Einstein. In Holland, for Leiden's academic elite Einstein was a particularly attractive candidate owing to his internationalist positions, which resonated strongly with their own efforts toward international reconciliation as inspired by van Vollenhoven's ideals and nationalist aspirations. Further, Leiden, as a neutral meeting ground, offered to Einstein the opportunity to act on his internationalist beliefs--along, of course, with the opportunity to interact with a group of first-rate physicists.

That, however, never really matured into a really longstanding, hands-on collaboration: Einstein never became a full-fledged participating member of the Leiden faculty. He would of course visit Leiden a fair number of times over the years, coauthor a few short papers with Ehrenfest, occasionally contribute to academic developments in Holland, and maintain a steady correspondence with Lorentz and Ehrenfest, which included many elaborate discussions on physics. Yet, by 1927 the Leiden University Fund had commuted Einstein's special professorship into a visiting professorship without any obligation to actually spend any time in Leiden. The remuneration would be set aside for any possible future visits of Einstein, but otherwise it would be spent on the education of young local physicists. This change in the terms of the professorship had been made at Einstein's request: the obligation to visit Leiden periodically had come to weigh too heavily on him, undoubtedly after many planned trips had been cancelled.[152] Einstein's special chair remained largely dormant until September 1952, when it finally was "retracted."[153]

The Dutch government had gone to great lengths to remain neutral in the Great War, and in its aftermath it very much feared to be drawn into the next maelstrom: it wished to keep out the revolutionary fever that was raging throughout Europe. Communists, especially following Troelstra's actions, therefore could not be appointed at Dutch state universities. Einstein, mistaken for his namesake, thus experienced an unusual delay in obtaining his special chair, since Carl Einstein was exactly the kind of radical that Dutch officials were most keen on keeping out of the Netherlands. Albert's confusion with Carl also helps to account for his broader reputation as a leftist radical, while his actual positions and persona were considerably more moderate. These need not have produced insuperable problems in Holland; indeed, in the end, they did not obstruct his appointment at Leiden.

Einstein's actual positions did produce problems in Germany. His difficulties in the new Weimar Republic were not with officials (he even would have easy access to some of its leading representatives), but with those who resisted relativity as revolutionary science. The confusion between and comparison of Albert and Carl Einstein has clearly shown that the resistance to relativity has to be seen, at least for some, as resistance to the November Revolution, the Weimar Republic, and in general the toppling of existing cultural and social values. For anti-relativists such as Paul Weyland and Ernst Gehrcke, both the old social order and what they regarded as proper scientific knowledge seemed to be at stake. Indeed, the vehemence of Einstein's opponents should be explained by their desire to resist marginalization--yet, they perceived their marginalization as taking place in the social realm at least as much as in the scientific realm. The result was that for Einstein's angry conservative opponents his revolution was just as unacceptable as Carl's.

In the midst of the German anti-relativity actions in the summer of 1920, Einstein told his friend and former collaborator, Marcel Grossmann, that:

> This world is a strange madhouse. Currently, every coachman and every waiter is debating whether relativity theory is correct. Belief in this matter depends on political party affiliation.[154]

When putting physics *circa* 1920 in perspective, we see that the history of the reception of relativity was strongly colored by the broader political and cultural contexts of the postwar period; in some circles, and not least in the public at large, it was almost entirely determined by these contexts. In particular, the history of the Leiden chair, and the confusions surrounding it, have shown that fears of revolution in the cultural as much as in the political domain, colored the initial perceptions of Albert Einstein, just as the confusions with Carl Einstein have made clear that the broad reception of relativity cannot be seen as separate from the revolutionary discourse that dominated Europe in the early years of the postwar period.

**Acknowledgments**
For insightful comments, I am grateful to David Baneke, A.J. Kox, Willem Otterspeer, Ze'ev Rosenkranz, and Milena Wazeck. I presented the earliest version of this article at a conference on Editing Centenary Scientific Manuscripts that was organized by Scott Walter of the Poincaré Archives in 2007 at the University of Nancy 2; I am grateful to Scott for inviting me to this conference. I am further indebted to my fellow editors of Volume 10 of *The Collected Papers of Albert Einstein* for a most rewarding collaboration, and to Ms. Toontje Jolles of the National Archive in The Hague for assistance. Finally, I thank Roger H. Stuewer for his meticulous and thoughtful editorial work on my paper.

**References**
Note: Translations from the German in Einstein's *Collected Papers* are mine.

For Einstein's statement on the disruptions of Nicolai's classes, see "In support of Georg Nicolai," *CPAE* 7, Doc. 32, pp. 282-283; 151.

[37] "Tumultszenen bei einer Einstein-Vorlesung" ("Uproar in the Lecture Hall"), *8-Uhr-Abendblatt* (February 13, 1920), *CPAE* 7, Doc. 33, pp. 284-288; 152.

[38] Otto J. de Jong, *Benoemingsbeleid aan de Rijksuniversiteiten (1876-1931). Rede bij de viering van de 364ste Dies Natalis der Rijksuniversiteit te Utrecht* (Utrecht: Utrecht University, 1982).

[39] *Ibid.*, p. 19.

[40] For an example see Peter Jan Knegtmans, "Professor Ernst Laqueur en de grenzen aan het internationalisme in de wetenschap in het interbellum," in Dorsman and Knegtmans, *Over de grens* (ref. 11), pp. 89-100.

[41] Peter Jan Knegtmans, "Politiek aan de Nederlandse universiteiten sedert 1876," in L.J. Dorsman and P.J. Knegtmans, ed., *Stille wijkplaatsen? Politiek aan de Nederlandse universiteiten sedert 1876* (Hilversum: Verloren, 2006), pp. 107-118, esp. pp. 109-110.

[42] Einstein to Elsa Einstein, May 9, 1920, *CPAE* 10, Doc. 9, pp. 252-253; 156.

[43] Elsa Einstein to Einstein, after May 9, 1920, *CPAE* 10, Doc. 10, pp. 253-255; 157-158, on 254; 157.

[44] Quoted from a letter to Hedwig and Max Born, January 27, 1920, in Rowe and Schulmann, *Einstein on Politics* (ref. 14), p. 410; see also pp. 406-408.

[45] Elsa Einstein to Einstein, May 24, 1920, *CPAE* 10, Doc. 30, pp. 275-276; 171-172.

[46] Einstein to Elsa Einstein May 20, 1920, *CPAE* 10, Doc. 22, pp. 267-277; 166.

[47] Einstein to Elsa Einstein, May 27, 1920, *CPAE* 10, Doc. 32, pp. 277-278; 173, on p. 277; 173.

[48] Lorentz to Einstein, May 27, 1920, *CPAE* 10, Doc. 35, p. 280; 174.

[49] De Visser to de Gijselaar and van Vollenhoven, March 22, 1920, NA.

[50] Parket van de Procureur Generaal to de Visser, March 19, 1920, NA.

[51] *Ibid*.

[52] Opperwachtmeester, Detachementscentrum, Koninklijke Marine, 1e Divisie to the Districts-Commandant in Breda, June 18, 1919, NA.

[53] Van Vollenhoven to de Visser, March 27, 1920, NA.

[54] De Visser to Minister van Buitenlandse Zaken, May 4, 1920, NA.

[55] Kamerlingh Onnes to de Visser, May 10, 1920, NA.

Siedler Verlag, 1989), pp. 382-383. However, not Carl, but Albert Einstein had been invited to this meeting; see Robert Schulmann, A.J. Kox, Michel Janssen, József Illy, and Karl von Meyenn, ed., *The Collected Papers of Albert Einstein.* Vol. 8. *The Berlin Years: Correspondence 1914-1918.* Part B. *1918* (Princeton: Princeton University Press, 1998), pp. 1029-1030; hereafter cited as *CPAE* 8. Furthermore, Vierset believed that Carl Einstein met on November 16 in Sint-Truiden with Adolphe Max, Mayor of Brussels, who was returning to the city from captivity in Germany; see Vierset, *Mes souvenirs* (ref. 98), pp. 491-492.

[103] Penkert, *Carl Einstein* (ref. 81), pp. 135-136.

[104] Troelstra, quoted in J.J. Woltjer, *Recent verleden: Nederland in de twintigste eeuw* (Amsterdam: Uitgeverij Balans, 2005), p. 147; on the SDAP, see pp. 146-160.

[105] The Dutch Cabinet's immediate responses, and its fear of Russian interventions, are well documented in the diaries of its Minister for Labor, P.J.M. Aalberse; see "Dagboek VI" and "Dagboek VII," available at the website of the *Huygens Instituut voor Nederlandse Geschiedenis*, <http://www.inghist.nl/Onderzoek/Projecten/Aalberse/Dagboeken>.

[106] J. de Bruijn, "Johannes Theodoor de Visser (1857-1932): Predikant, politicus en onderwijshervormer," in Paul E. Werkman and Rolf E. van der Woude, ed., *Wie in de politiek gaat, is weg? Protestantse politici en de christelijk-sociale beweging* (Hilversum: Verloren, 2009), pp. 93-122, on pp. 101-102.

[107] Ehrenfest to Einstein, April 13, 1920, *CPAE* 9, Doc. 373, pp. 500-505; 309-313, on p. 503; 312.

[108] David Baneke, "'Hij kan toch moeilijk de sterren in de war schoppen.' De afwijzing van Pannekoek als adjunct-direkteur van de Leidse Sterrewacht in 1919," *Gewina. Tijdschrift voor de Geschiedenis der Geneeskunde, Natuurwetenschappen, Wiskunde en Techniek* **27** (2004), 1-13; see also David Baneke, "Teach and Travel: Leiden Observatory and the Renaissance of Dutch Astronomy in the Interwar Years," *Journal for the History of Astronomy* **41** (2010), 167-198, esp. 170-173.

[109] E.P.J. van den Heuvel, "Antonie Pannekoek (1873-1960). Socialist en sterrenkundige," in J.C.H. Blom, P.H.D. Leupen, P. de Rooy, T.J. Veen, and L. Kooijmans, ed., *Een brandpunt van geleerdheid in de hoofdstad. De Universiteit van Amsterdam rond 1900 in vijftien portretten* (Hilversum: Verloren, 1992), pp. 229-245. For completeness, it should be noted that Lenin wrote a pamphlet against Pannekoek's views, after Pannekoek had distanced himself from Lenin's revolution.

[110] De Sitter and Pannekoek, cited in Baneke "'Hij kan toch moeilijk de sterren'" (ref. 108), p. 9.

[111] De Visser to the curators of Leiden University, May 3, 1919, cited in *ibid.*, p. 9.

[112] De Gijselaar to de Sitter, May 16, 1919, cited in *ibid.*, p. 10.

[113] Motions of the Tweede Kamer, cited in *ibid.*, p. 11.

[114] De Visser, in the Motions of the Eerste Kamer, cited in *ibid.*, p. 11.